\documentclass[amsmath,amssymb,floatfix,lengthcheck,preprintnumbers,prd,showpacs,superscriptaddress,nofootinbib,twocolumn,aps]{revtex4-1}
\usepackage{graphicx}
\usepackage{slashed}
\usepackage{xcolor}
\usepackage{mathtools}
\newcommand{\beq}{\begin{equation}}
\newcommand{\eeq}{\end{equation}}
\newcommand{\beqs}{\begin{eqnarray}}
\newcommand{\eeqs}{\end{eqnarray}}
\newcommand{\lsim}{\mathrel{\raisebox{-.6ex}{$\stackrel{\textstyle<}{\sim}$}}}
\newcommand{\gsim}{\mathrel{\raisebox{-.6ex}{$\stackrel{\textstyle>}{\sim}$}}}

\newcommand{\spinvec}[2]{\left(\!\begin{array}{c}#1 \\ #2\end{array} \!\right)}
\newcommand{\spinmat}[4]{\left(\!\begin{array}{cc}#1&#2\\#3&#4\end{array} \!\right)}

\long\def\/*#1*/{}
\definecolor{red}{rgb}{1.0, 0, 0}

\newcommand{\ra}{\rightarrow}
\newcommand{\eps}{\epsilon}

\begin{document}

\title{Stealth Dark Matter:  Dark scalar baryons through the Higgs portal}

\author{T.~Appelquist}
\affiliation{Department of Physics, Sloane Laboratory, Yale University,
             New Haven, Connecticut 06520, USA}

\author{R.~C.~Brower}
\affiliation{Department of Physics, Boston University,
	Boston, Massachusetts 02215, USA}
\author{M.~I.~Buchoff}
\affiliation{Institute for Nuclear Theory, Box 351550, Seattle, WA 98195-1550, USA}

\author{G.~T.~Fleming}
\affiliation{Department of Physics, Sloane Laboratory, Yale University,
             New Haven, Connecticut 06520, USA}
\author{X.-Y.~Jin}
\affiliation{Argonne Leadership Computing Facility, Argonne National Laboratory, Argonne, IL 60439, USA}
\author{J.~Kiskis}
\affiliation{Department of Physics, University of California,
	Davis, California 95616, USA}
\author{G.~D.~Kribs}
\affiliation{Department of Physics, University of Oregon, Eugene, OR, 97403 USA}

\author{E.~T.~Neil}
\affiliation{Department of Physics,
        University of Colorado, Boulder, CO 80309, USA}
\affiliation{RIKEN-BNL Research Center, Brookhaven National Laboratory, Upton, NY 11973, USA}
\author{J.~C.~Osborn}
\affiliation{Argonne Leadership Computing Facility, Argonne National Laboratory, Argonne, IL 60439, USA}
\author{C.~Rebbi}
\affiliation{Department of Physics, Boston University,
	Boston, Massachusetts 02215, USA}
\author{E.~Rinaldi}
\affiliation{Lawrence Livermore National Laboratory, Livermore, California 94550, USA}
\author{D.~Schaich}
\affiliation{Department of Physics, Syracuse University, Syracuse, NY 13244, USA}
\author{C.~Schroeder}
\affiliation{Lawrence Livermore National Laboratory, Livermore, California 94550, USA}
\author{S.~Syritsyn}
\affiliation{RIKEN-BNL Research Center, Brookhaven National Laboratory, Upton, NY 11973, USA}
\author{P.~Vranas}
\affiliation{Lawrence Livermore National Laboratory, Livermore, California 94550, USA}
\author{E.~Weinberg}
\affiliation{Department of Physics, Boston University,
	Boston, Massachusetts 02215, USA}
\author{O.~Witzel}
\altaffiliation[Present address: ]{Higgs Centre for Theoretical Physics, School of Physics \& Astronomy, The University of Edinburgh, EH9 3FD, UK}
\affiliation{Center for Computational Science, Boston University, Boston, MA, USA}

\collaboration{Lattice Strong Dynamics (LSD) Collaboration}

\begin{abstract}

We present a new model of ``Stealth Dark Matter'': 
a composite baryonic scalar of an $SU(N_D)$ strongly-coupled theory 
with even $N_D \geq 4$.  All mass scales 
are technically natural, and dark matter stability is automatic
without imposing an additional discrete or global symmetry. 
Constituent fermions transform in vector-like 
representations of the electroweak group that permit both 
electroweak-breaking and electroweak-preserving mass terms.
This gives a tunable coupling of stealth
dark matter to the Higgs boson independent of the 
dark matter mass itself.  We specialize to $SU(4)$, and 
investigate the constraints on the
model from dark meson decay, electroweak precision 
measurements, basic collider limits, and spin-independent direct 
detection scattering through Higgs exchange.  We exploit our earlier
lattice simulations that determined the composite spectrum 
as well as the effective Higgs coupling of stealth dark matter in order 
to place bounds
from direct detection, excluding constituent fermions with dominantly
electroweak-breaking masses.
A lower bound on the dark baryon mass $m_B \gsim 300$~GeV 
is obtained from the indirect requirement that the lightest dark meson 
not be observable at LEP II\@.  We briefly survey some intriguing 
properties of stealth dark matter that are worthy of future study, 
including: collider studies of dark meson production and decay;
indirect detection signals from annihilation; 
relic abundance estimates for both symmetric and asymmetric mechanisms;
and direct detection through electromagnetic polarizability, 
a detailed study of which will appear in a companion paper.

\end{abstract}

\preprint{INT-PUB-15-004, 
LLNL-JRNL-667446}

\maketitle

\section{\textbf{Introduction}}

Composite dark matter, made up of electroweak-charged constituents
provides a straightforward mechanism for obtaining viable electrically-neutral
particle dark matter that can yield the correct cosmological abundance 
while surviving direct and indirect detection search limits, e.g., 
\cite{Aprile:2012nq,Agnese:2013rvf,Akerib:2013tjd}.
In this paradigm, the dark sector consists of fermions that 
transform under the electroweak group and a new, strongly-coupled 
non-Abelian dark force.  This was considered long ago in the
context of technicolor theories, where the strong dynamics was
doing double-duty to both break electroweak symmetry and provide
a dark matter candidate
\cite{Nussinov:1985xr,Chivukula:1989qb,Barr:1990ca,Kaplan:1991ah}.

In this paper, electroweak symmetry breaking is accomplished
through the weakly-coupled Standard Model Higgs mechanism, 
while the new strongly-coupled sector is reserved solely
for providing a viable dark matter candidate.
This dark sector is not easy to detect in dark matter 
detection experiments or in collider experiments,
and so we give it the name ``Stealth Dark Matter''. 
Earlier work in this direction includes
\cite{Nussinov:1985xr,Chivukula:1989qb,Barr:1990ca,Barr:1991qn,Kaplan:1991ah,Chivukula:1992pn,Bagnasco:1993st,Pospelov:2000bq,Dietrich:2006cm,Foadi:2008qv,Khlopov:2008ty,Mardon:2009gw,Kribs:2009fy,Lisanti:2009am,Alves:2010dd,Khlopov:2010pq,Buckley:2012ky,Cline:2013zca,Heikinheimo:2013fta, Heikinheimo:2014xza, Yamanaka:2014pva, Hochberg:2014dra,Boddy:2014yra,Boddy:2014qxa,Hochberg:2014kqa}, and except for \cite{Lewis:2011zb,Hietanen:2012sz,Appelquist:2013ms, Hietanen:2013fya,Appelquist:2014jch,Detmold:2014qqa,Detmold:2014kba,Fodor:2015eea}, was often 
limited by the inability to perturbatively calculate the 
spectrum and form factors due to strong coupling.    

The proposed dark matter candidate is
a scalar baryon of $SU(N_D)$, and hence $N_D$ must be even.\footnote{Fermionic baryons arising from odd $N_D$ were considered in Ref.~\cite{Appelquist:2013ms}
where the limit $M \gsim 10$~TeV was found to avoid
the direct detection constraints from the magnetic dipole interaction.}
We take the dark fermions to be in vector-like 
representations of the electroweak group. 
Hence, the constituent dark fermions can acquire 
bare mass terms (fermion masses that do not require 
electroweak symmetry breaking) while also permitting Yukawa 
interactions that marry dark fermion electroweak doublets with singlets.
This yields a theory in which dark matter couples to the Higgs boson 
in a tunable way that is essentially independent of the 
dark matter mass itself.  This is somewhat analogous to dark sector
models with a dark U(1) portal 
(e.g.,~\cite{Pospelov:2007mp,ArkaniHamed:2008qn,Pospelov:2008zw,Cheung:2009qd,Morrissey:2009ur}), where the coupling to the 
Standard Model is tunable through an otherwise arbitrary parameter -- 
the kinetic mixing between the dark U(1) and hypercharge.

The existence of both electroweak-preserving and electroweak-breaking
masses for the dark fermions provides two main benefits. 
First, given that the Higgs boson couples electroweak doublets 
with singlets, the global flavor symmetries of the dark fermions 
can be completely broken to just dark baryon number.
All mesons can decay through an 
electroweak process (e.g., electrically charged mesons through 
$W$ exchange) or through the usual chiral anomaly 
(e.g., the lightest neutral meson). 
Ensuring that these particles decay before big-bang nucleosynthesis
sets a weak lower bound on the Higgs interaction strength.
(This is in contrast with \cite{Kilic:2009mi,Buckley:2012ky} 
where additional interactions were required to ensure mesons decay, 
e.g., through higher dimensional operators). 
The second reason is related to the orientation of the chiral
condensate after the dark force confines.  Large vector-like masses
for the dark fermions ensure that the condensate can be aligned 
toward the electroweak-preserving direction, 
and thus the dark sector leads to only small corrections to 
electroweak precision measurements.  We estimate the size of
these corrections in this paper. 

There are many appealing features of an electroweak-neutral composite dark matter
candidate made up from fermions transforming under the electroweak group,
including:
\begin{itemize}
\item All of the dimensionful scales are technically natural, 
since they arise from fermion masses (vector-like and electroweak breaking)
and the confinement of a strong-coupled dark force. 
\item Dark matter stability is an automatic consequence of 
dark baryon number conservation.  No additional global discrete or 
continuous symmetries are required.  
For $N_D \ge 3$, operators involving dark baryon decay are 
necessarily dimension-6 or higher, and thus safe from
GUT-scale or Planck-scale suppressed violations of dark baryon number.
\item There are no dimension-4 
interactions of the composite dark matter particle 
with the Standard Model except with the Higgs boson. 
The direct detection scattering cross section is thus 
automatically suppressed compared with a generic
elementary WIMP candidate.
\item Higher dimensional interactions of the dark matter
with the Standard Model are suppressed, in the nonrelativistic limit, 
by several powers of the dark matter mass.  
For a composite scalar, the leading operators 
are charge radius (at dimension-6) and polarizability (at dimension-7). 
The impact of these (and other) operators on the dark matter scattering 
cross section in direct detection experiments has been studied in
\cite{Chivukula:1992pn,Bagnasco:1993st,Pospelov:2000bq,Sigurdson:2004zp,Gudnason:2006ug,Alves:2009nf,Kribs:2009fy,Barbieri:2010mn,Banks:2010eh,Chang:2010en,Barger:2010gv,Weiner:2012cb}.
\item Interactions of the dark baryon through the neutral weak current, 
the charge radius interaction, as well as the contributions 
to the electroweak precision $T$ parameter, are simultaneously
eliminated if the fermion interactions obey a global 
custodial $SU(2)$ symmetry.  Additionally, as we will see,
dark matter electric neutrality also follows from custodial $SU(2)$.
To simplify our analysis, here we will primarily study the subset of stealth dark matter parameter space
in which the custodial $SU(2)$ is preserved.
(This simplification is very familiar from composite Higgs 
theories, e.g.~\cite{Agashe:2003zs}). 
\item The abundance of a strongly-coupled dark scalar baryon 
could arise through several mechanisms:  an asymmetric abundance
(such as through electroweak sphalerons \cite{Barr:1990ca,Barr:1991qn}
or other mechanisms \cite{Zurek:2013wia}), when the mass
is not too large $\lsim$~few TeV, or a symmetric abundance, 
when the mass is large (perhaps $\sim \mathcal{O}(100)$~TeV) 
\cite{Griest:1989wd,Buckley:2012ky,Nussinov:2014qva}. 
\end{itemize}

We focus mainly on a confining
$SU(4)$ gauge theory dark sector with dark fermions transforming 
non-trivially under the electroweak group.  We apply our 
recent results \cite{Appelquist:2014jch} using lattice simulations for 
the spectrum and effective Higgs interaction for $SU(4)$.
As emphasized in \cite{Appelquist:2013ms,Appelquist:2014jch},
this theory is well suited for lattice calculations
since we are not interested in the chiral limit of 
vanishing dark fermion masses.  Indeed, lattice simulations can
efficiently simulate the parameter region where the dark fermion 
masses are comparable to the confinement scale, exactly where the 
perturbative estimates are least useful.  

The organization of the paper is as follows.  
In Sec.~\ref{sec:constructing} we discuss the assumptions
and requirements to construct our stealth dark matter model.
In Sec.~\ref{sec:darkfermioninteractions} we detail the
dark fermion interactions and masses.  
In addition, 
we write the electroweak
currents in terms of the dark fermion mass eigenstates of 
the theory, detailed in Appendix~\ref{sec:weakcurrents}.  
Until this point, the discussion of the model
is general.  In Sec.~\ref{sec:simplifications}, 
we simplify the parameter
space for phenomenological and calculational purposes,
applying a global custodial $SU(2)$ symmetry and taking the 
approximately symmetric dark fermion mass matrix limit.
Then in Sec.~\ref{sec:meson} we discuss the light non-singlet mesons 
in the theory, in particular their decay rates and constraints 
from non-observation at LEP II\@.  In Sec.~\ref{sec:pew} we
discuss the stealth dark matter contributions to the 
$S$ parameter, and demonstrate the parametric suppression
that happens in several regimes.  In Sec.~\ref{sec:fermionhiggs}
we obtain the Higgs boson coupling to the dark fermions.
Then in Sec.~\ref{sec:boundshiggs} we apply our previous
model-independent results on the $SU(4)$ spectrum and 
effective Higgs coupling to stealth dark matter.
We obtain the bounds on the parameter space from
the non-observation of a spin-independent direct detection 
signal at LUX\@.  We briefly discuss the relic abundance
of stealth dark matter in Sec.~\ref{sec:abundance}.  
Finally we conclude with a discussion in Sec.~\ref{sec:discussion}.

\section{\textbf{Constructing a viable model}}
\label{sec:constructing}

\subsection{Basic assumptions}

We assume that the dark matter candidate is a 
composite particle of a non-Abelian, confining gauge theory 
based on the group $SU(N_{D})$ with $N_f$ flavors of fermions transforming in the 
fundamental representation. 
The number $N_f$ is restricted by only the condition of confinement. 
For reasons outlined in the introduction (abundance, detectability),
the dark fermions carry electroweak charges. Our model includes 
a tunable Higgs ``portal'' coupling between the dark sector and 
the Standard Model via dimension-4 Higgs couplings.\footnote{Other 
portals, such as a dark gauged U(1) group 
that kinetically mixes with hypercharge, are neither present 
nor required here.}
We do not consider
QCD-colored dark fermions since with $N_D \neq 3$, dark baryons 
would not generally be color singlets.\footnote{The obvious exception, 
when $N_D = N_c = 3$, is discussed in \cite{Bai:2013xga, Appelquist:2013ms}, 
which is not a focus for us due to the baryons
being fermions.  Construction of QCD-singlet dark baryons with $N_D=6,12,18,...$ may be possible, but
we do not study this possibility further here.}

\subsection{Requirements}

We require dark matter stability to be automatic, 
arising from a global symmetry.  This motivates 
considering the dark baryon of the non-Abelian dark sector 
to be the dark matter \cite{Nussinov:1985xr,Chivukula:1989qb,Barr:1990ca}.
In the presence of GUT-scale
or Planck-scale suppressed operators, the stability of the dark baryon 
should be sufficient to avoid cosmological constraints.  

The requirement of a sufficiently preserved accidental baryon number 
disfavors a dark $SU(2)$ group.  First, there is no automatic 
baryon number in $SU(2)$ because there is no fundamental distinction 
between mesons and baryons.  Imposing a global $U(1)$ baryon number 
is possible (e.g.\ see \cite{Kribs:2009fy}) 
but in addition baryon number violating dimension-5 
Planck-suppressed operators such as 
$f_{\rm dark} f_{\rm dark} H^{\dagger} H/M_{\rm Pl}$ must be absent, where $f_{\rm dark}$ is the dark fermion.  
(Otherwise, the dark $SU(2)$ baryon would decay on 
a timescale much shorter than the age of the Universe.)

For $N_D \ge 3$, operators involving dark baryon decay are 
necessarily dimension-6 or higher, and thus safe from
GUT-scale or Planck-scale suppressed violations of dark baryon number.
$SU(N_D)$ with odd $N_D$ is a perfectly interesting theory, 
having been studied before for $N_D = 3$ by our collaboration \cite{Appelquist:2013ms}.
There it was found
that a fermionic dark baryon has a magnetic dipole interaction 
that leads to a significant contribution to spin-independent scattering.  
Constraints from the XENON100 experiment were satisfied only when
the dark matter mass $M \gsim 10$~TeV \cite{Appelquist:2013ms}.
This strong constraint on the mass scale implies the model is
difficult to test at near-future colliders.

The magnetic dipole interaction (and other higher dimensional operators 
that require spin) are absent when the dark baryon is a scalar.  
We are thus naturally led to $SU(N_D)$ with even $N_D \ge 4$, for which 
the otherwise strong constraints from direct detection are weakened,
lowering the scales of interest into a regime that
can be probed by colliders and other detection strategies.

We assume the dark fermions have masses $M_f$ on the order of the
$SU(N_D)$ confinement scale $\Lambda_D$. If the masses were much
smaller, the dark sector would 
contain light pseudo-Goldstone pions that transform under the 
electroweak group, which are strongly constrained by collider 
experiments.  
A dark sector with purely vector-like fermion masses has 
approximately stable electrically-charged mesons due to  
dark flavor symmetries.
Conversely, a dark sector with purely electroweak breaking
fermion masses has a dark matter candidate that is ruled out 
by spin-independent direct detection through single Higgs exchange.  
(For example, quirky dark matter \cite{Kribs:2009fy} 
is now completely ruled out by 
Higgs exchange, given the direct detection bounds from 
LUX \cite{Akerib:2013tjd}
combined with the relatively light Higgs mass 
\cite{Aad:2012tfa,Chatrchyan:2012ufa}.)
Fermions with both vector-like and (small) electroweak breaking 
contributions to their masses can avoid both problems.

We require the lightest dark baryon to be electrically neutral. 
We also require Higgs couplings at dimension-4 to pairs of 
dark fermions.  These two requirements impose restrictions on 
the electroweak charges of the dark fermions.  

One solution is familiar from old technicolor theories 
(e.g.~\cite{Peskin:1980gc,Preskill:1980mz}): requiring the 
dark fermion charges to roughly satisfy $|Y| \lsim |T_3|$
where $T_3$ is the $SU(2)_L$ isospin.  Choosing doublets ($|T_3| = 1/2$ 
under $SU(2)_L$) then gives a finite number of discrete possibilities.

A simple model that satisfies all of these requirements 
is shown in Table~\ref{tab:particles}.  The electric charges of the 
dark fermions in the broken electroweak phase are $Q = \pm 1/2$, 
ensuring all hadrons have integer electric charges.
So long as the lightest $Q = 1/2$ and $Q = -1/2$ dark fermions
are close in mass, the lightest baryon will be a scalar and
electrically neutral.  Finally, with the assignments shown in
Table~\ref{tab:particles},
all gauge (and global) anomalies vanish, 
which is automatic with fermions that transform under vector-like 
representations of the $SU(N_D)$ and electroweak groups.

\section{\textbf{Dark fermion interactions and masses}}
\label{sec:darkfermioninteractions}

\begin{table}[t]
\renewcommand{\arraystretch}{1.3}
\begin{center}
\begin{tabular}{c|c|c|c}
Field & $SU(N_D)$             & ($SU(2)_{L}$, $Y$)        & $Q$ \\ \hline
$F_1 = \spinvec{F^u_1}{F^d_1}$ & $\mathbf{N}$            
  & $(\mathbf{2}, 0)$         & $\spinvec{+1/2}{-1/2}$ \\
$F_2 = \spinvec{F^u_2}{F^d_2}$ & $\overline{\mathbf{N}}$ 
  & $(\mathbf{2}, 0)$   & $\spinvec{+1/2}{-1/2}$ \\
$F_3^u$ & $\mathbf{N}$  
  & $(\mathbf{1}, +1/2)$  & $+1/2$ \\
$F_3^d$ & $\mathbf{N}$  
  & $(\mathbf{1}, -1/2)$  & $-1/2$ \\
$F_4^u$ & $\overline{\mathbf{N}}$  
  & $(\mathbf{1}, +1/2)$  & $+1/2$ \\
$F_4^d$ & $\overline{\mathbf{N}}$  
  & $(\mathbf{1}, -1/2)$  & $-1/2$ 
\end{tabular}
\end{center}
\caption{Dark fermion particle content of the stealth dark matter model.
All fields are two-component (Weyl) spinors.  $SU(2)_L$ refers to the 
Standard Model electroweak gauge group, and $Y$ is the hypercharge.
In the broken phase of the electroweak theory, the dark fermions
have the electric charge $Q = T_3 + Y$ as shown.}

\label{tab:particles}
\end{table}

The fermions $F_i^{u,d}$ transform under
a global $U(4) \times U(4)$ flavor symmetry with 
[$SU(2) \times U(1)$]$^4$ 
surviving after the weak gauging of the electroweak
symmetry.  From this large global symmetry, one $SU(2)$ 
(diagonal) subgroup will be identified with $SU(2)_L$, 
one $U(1)$ subgroup will be identified with $U(1)_Y$, 
and one $U(1)$ will be identified with dark baryon number. 
The total fermionic content of the model is therefore 
8 Weyl fermions that pair up to become 4 Dirac fermions 
in the fundamental or anti-fundamental representation 
of $SU(N_D)$ with electric charges of 
$Q \equiv T_{3,L} + Y = \pm 1/2$. 
We use the notation where the superscript $u$ or $d$ 
(as in $F^u$, $F^d$ and later $\psi^u$, $\psi^d$, $\Psi^u$, $\Psi^d$)
denotes a fermion with electric charge of $Q=1/2$ or $Q=-1/2$ 
respectively. 

The fermion kinetic terms in the Lagrangian are given by 
\begin{equation}
\mathcal{L} \; \supset \; 
  \sum_{i=1,2} i F_i^\dagger \bar{\sigma}^\mu D_{i,\mu} F_i 
  + \sum_{i=3,4; j=u,d} i {F_i^j}^\dagger \bar{\sigma}^\mu D_{i,\mu}^j F_i^j
  \, ,
\end{equation}
where the covariant derivatives are
\begin{eqnarray}
D_{1,\mu} &\equiv& \partial_\mu - i g W^a_\mu \sigma^a /2 
                                - i g_D G^b_\mu t^b \\ 
D_{2,\mu} &\equiv& \partial_\mu - i g W^a_\mu \sigma^a /2 
                                + i g_D G^b_\mu {t^b}^* \\ 
D_{3,\mu}^j &\equiv& \partial_\mu - i g' Y^j B_\mu 
                                  - i g_D G^b_\mu t^b \\
D_{4,\mu}^j &\equiv& \partial_\mu - i g' Y^j B_\mu 
                                  + i g_D G^b_\mu {t^b}^* 
\end{eqnarray}
with the interactions among the electroweak group
and the new $SU(N_D)$.  Here $Y^u = 1/2$, $Y^d = -1/2$ and 
$t^b$ are the representation matrices for the fundamental 
of $SU(N_D)$. 

The vector-like mass terms allowed by the gauge symmetries are
\begin{equation}
\mathcal{L} \supset M_{12} \epsilon_{ij} F_1^i F_2^j 
  - M_{34}^u F_3^u F_4^d + M_{34}^d F_3^d F_4^u + h.c., 
\end{equation}
where $\epsilon_{12} \equiv \epsilon_{ud} = -1 = -\epsilon^{12}$
and the relative minus signs between 
the mass terms have been chosen for later convenience.  
The mass term $M_{12}$ explicitly breaks an [$SU(2) \times U(1)$]$^2$ 
global symmetry down to the diagonal $SU(2)_{\rm diag} \times U(1)$
where the $SU(2)_{\rm diag}$ is identified with $SU(2)_L$.  
The mass terms $M_{34}^{u,d}$ explicitly break the remaining 
[$SU(2) \times U(1)$]$^2$ down to $U(1) \times U(1)$ where 
one of the $U(1)$'s is identified with $U(1)_Y$.  
(In the special case when $M_{34}^{u} = M_{34}^{d}$, 
the global symmetry is enhanced to 
$SU(2) \times U(1)$, where the global $SU(2)$ acts 
as a custodial symmetry.) 
Thus, after weakly gauging the electroweak
symmetry and writing arbitrary vector-like mass terms, 
the unbroken flavor symmetry is $U(1) \times U(1)$.

Electroweak symmetry breaking mass terms arise from coupling to 
the Higgs field $H$ that we take to be in the 
$(\mathbf{2}, +1/2)$ representation.  They are given by
\begin{eqnarray}
\mathcal{L} &\supset& y_{14}^u \epsilon_{ij} F_1^i H^j F_4^d 
                      + y_{14}^d F_1 \cdot H^\dagger F_4^u      \nonumber \\
 & &{}                - y_{23}^d \epsilon_{ij} F_2^i H^j F_3^d 
                      - y_{23}^u F_2 \cdot H^\dagger F_3^u 
                      + h.c. \, ,
\label{eq:higgsterms}
\end{eqnarray}
where again the relative minus signs are chosen for later convenience.
After electroweak symmetry breaking, $H = ( 0 \;\; v/\sqrt{2} )^T$, 
with $v \simeq 246$~GeV\@. 
Replacing the Higgs field by its VEV in Eq.~(\ref{eq:higgsterms}), 
we obtain mass terms
for the fermions, in 2-component notation,
\begin{eqnarray} 
\mathcal{L} &\supset& - (F_1^u \;\; F_3^u) M^u \spinvec{F_2^d}{F_4^d} 
  - (F_1^d \;\; F_3^d) M^d \spinvec{F_2^u}{F_4^u} \nonumber \\
    & &{} + h.c. \, ,
\label{eq:2compmassmatrix}
\end{eqnarray}
with the mass matrices given by 
\begin{eqnarray}
M^u &\equiv& 
  \spinmat{M_{12}}{y_{14}^u v/\sqrt{2}}{y_{23}^u v/\sqrt{2}}{M_{34}^u} 
  \label{eq:upmassmatrix} \\ 
M^d &\equiv& -
  \spinmat{M_{12}}{y_{14}^d v/\sqrt{2}}{y_{23}^d v/\sqrt{2}}{M_{34}^d} \, .
  \label{eq:downmassmatrix} 
\end{eqnarray}

These Yukawa couplings break the remaining 
$U(1) \times U(1)$ flavor symmetry to $U(1)_D$ dark baryon number.
The mass matrices $M^u$ and $M^d$ correspond to the masses of 
two sets of fermions with electric charge $Q = +1/2$ and $Q = -1/2$ 
respectively, in the fundamental representation of $SU(N_D)$. 
The two biunitary mass matrices can be diagonalized by four independent 
rotation angles
\begin{eqnarray}
\left( \begin{array}{cc} M^u_1 & 0 \\ 0 & M^u_2 \end{array} \right) 
&=& R(\theta^u_1)^{-1} M^u R(\theta^u_2) 
  \label{eq:diagu} \\
\left( \begin{array}{cc} M^d_1 & 0 \\ 0 & M^d_2 \end{array} \right) 
&=& R(\theta^d_1)^{-1} M^d R(\theta^d_2) \, ,
  \label{eq:diagd}
\end{eqnarray}
where the rotation matrices are defined by
\begin{eqnarray}
R(\theta_i^j) &\equiv& 
   \spinmat{\cos\theta_i^j}{-\sin\theta_i^j}{\sin\theta_i^j}{\cos\theta_i^j} 
    \, .
\end{eqnarray}
The 2-component mass eigenstate spinors are thus 
\begin{eqnarray}
\spinvec{\psi_1^u}{\psi_2^u} &=& R(\theta^u_1) \spinvec{F_1^u}{F_3^u} 
   \label{eq:vec1} \\
\spinvec{\psi_1^d}{\psi_2^d} &=& R(\theta^u_2) \spinvec{F_2^d}{F_4^d} 
   \label{eq:vec2} \\
\spinvec{\chi_1^d}{\chi_2^d} &=& i R(\theta^d_1) \spinvec{F_1^d}{F_3^d} 
   \label{eq:vec3} \\
\spinvec{\chi_1^u}{\chi_2^u} &=& i R(\theta^d_2) \spinvec{F_2^u}{F_4^u} \, ,
   \label{eq:vec4} 
\end{eqnarray}
where the extra phase in Eqs.~(\ref{eq:vec3}),(\ref{eq:vec4}) 
ensures the $Q=-1/2$ fermions will have positive mass eigenvalues.

The Lagrangian for the fermion mass eigenstates becomes
\begin{eqnarray}
\mathcal{L} &\supset& - \sum_{i=1}^2 \left( M_i^u \psi_i^u \psi_i^d 
            + M_i^d \chi_{i}^d \chi_{i}^u + h.c. \right) 
\end{eqnarray}
where the mass eigenvalues are $M_{1,2}^u$ for $Q=1/2$,
and the distinction between fermions $\psi$ and $\chi$ allows us to write 
the $Q=-1/2$ fermion masses as $M_{1,2}^d$. 
The Dirac spinor mass eigenstates are constructed from the 
2-component Weyl spinor mass eigenstates in the usual way,
\begin{eqnarray}
\Psi_i^u &\equiv& \spinvec{\psi^u_i}{{\psi^d_i}^\dagger} \qquad i=1,2 
  \label{eq:diracup} \\
\Psi_i^d &\equiv& \spinvec{\chi^d_{i}}{{\chi^u_{i}}^\dagger} \qquad i=1,2
  \label{eq:diracdown} 
\end{eqnarray}
giving the Dirac fermion masses
\begin{eqnarray}
\mathcal{L} &\supset& 
  - \sum_{i=1}^2 \left(   M_i^u \overline{\Psi}_i^u \Psi_i^u
                        + M_i^d \overline{\Psi}_i^d \Psi_i^d \right) \, .
\end{eqnarray}

The fermion masses themselves are obtained from a straightforward
diagonalization of the mass matrices, 
\begin{equation}
M^u_{1,2} = \frac{M_{12} + M_{34}^u}{2} \mp 
\left[ \left( \frac{M_{12} - M_{34}^u}{2} \right)^2 
       + \frac{y^u_{14} y^u_{23} v^2}{2} \right]^{1/2} \, ,
\end{equation}
with mixing angles
\begin{equation}
\tan 2\theta_1^u = 
\frac{2 \sqrt{2} v (M_{12} y^u_{23} + M^u_{34} y^u_{14} )}{
2 M_{12}^2 - 2 (M_{34}^u)^2 + (y^u_{14} v)^2 - (y^u_{23} v)^2} 
\end{equation}
\begin{equation}
\tan 2\theta_2^u = 
\frac{2 \sqrt{2} v (M_{12} y^u_{14} + M^u_{34} y^u_{23} )}{
2 M_{12}^2 - 2 (M_{34}^u)^2  -(y^u_{14} v)^2  + (y^u_{23} v)^2}
\, ,
\end{equation}
with identical expressions for $M^d_{1,2}$ and $\tan 2\theta^d_{1,2}$
with the replacement $u \leftrightarrow d$ everywhere.  

It is important to note that the electroweak currents ($j^\mu_{+}$, $j^\mu_{-}$, $j^\mu_{3}$, $j^\mu_{Y}$) play an important role in the upcoming phenomenological discussions.  Due to the extended expressions for these quantities in terms of our Dirac spinors, we have relegated a detailed derivation of the electroweak currents to Appendix~\ref{sec:weakcurrents}. 

\section{\textbf{Simplifications}}
\label{sec:simplifications}

Our main interest is the more specialized case where 
the lightest $Q = +1/2$ and $Q = -1/2$ fermions are 
degenerate in mass to a very good approximation. 
This leads to a neutral scalar baryon with a vanishing charge radius.
While there are several ways this could be accomplished, we can simply
impose a custodial $SU(2)$ global symmetry on the Lagrangian.   In order to simplify notation, we define  $c_i^j \equiv \cos\theta_i^j$, $s_i^j \equiv \sin\theta_i^j$ and $P_{L,R} = (1 \mp \gamma_5)/2$.  In the custodial $SU(2)$ symmetric theory, $c_i^u=c_i^d$ and $s_i^u=s_i^d$.

\subsection{Custodial SU(2)}

\begin{figure}[t]
\begin{center}
\includegraphics[width=0.49\textwidth]{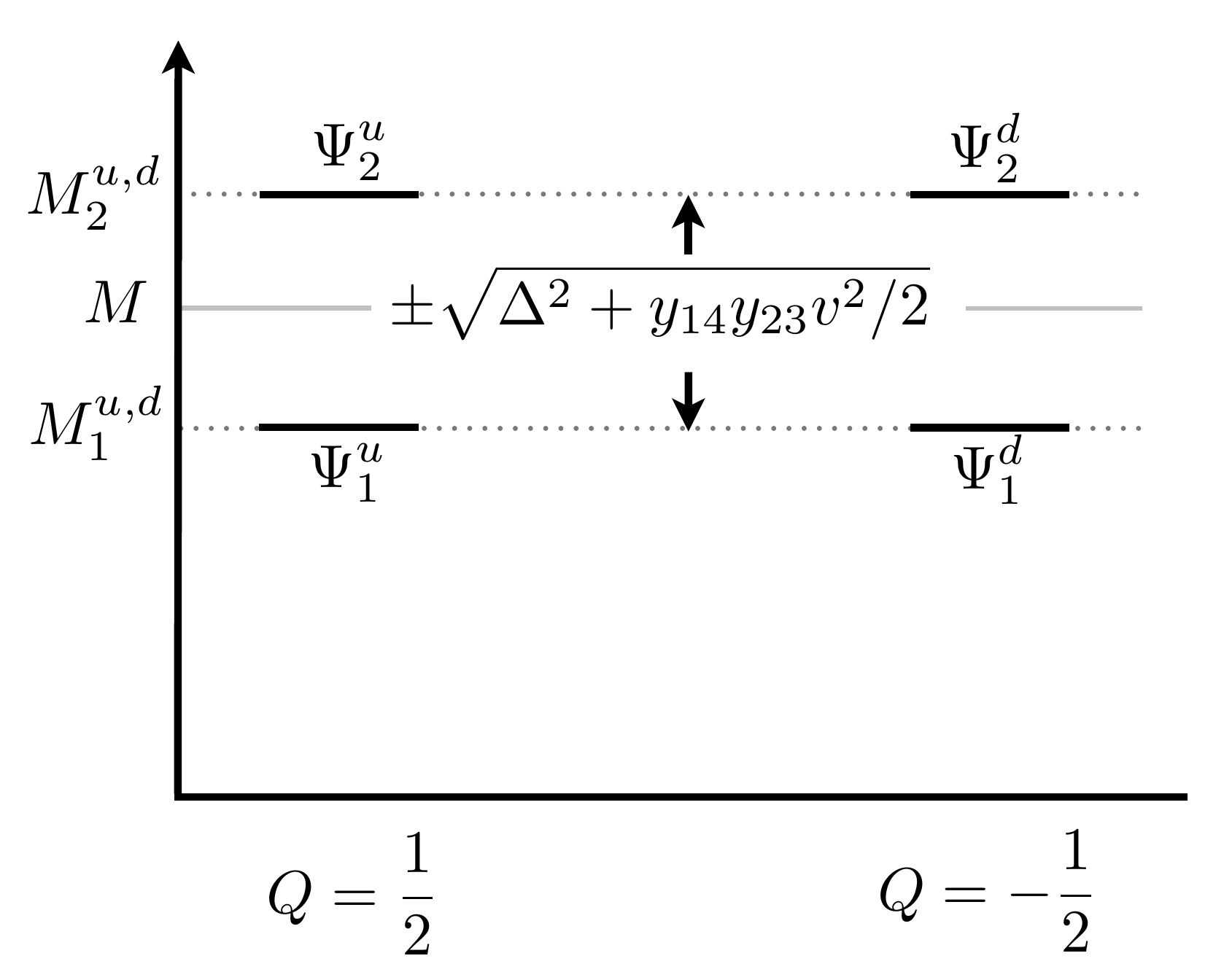}
\end{center}
\caption{Illustration of the fermion mass spectra considered
in the paper.  Four Dirac fermions 
($\Psi^u_1$, $\Psi^d_1$, $\Psi^u_2$, $\Psi^d_2$) 
have masses 
($M_1^u$, $M_1^d$, $M_2^u$, $M_2^d$). 
The $u$ ($d$) fermions
have electric charge $Q = +1/2$ ($Q = -1/2$); we assume an exact 
custodial $SU(2)$ global symmetry that ensures each $Q = +1/2$ fermion is 
accompanied by a $Q = -1/2$ fermion with equal mass as shown in
the figure.  If $\Delta \ll \sqrt{y_{14} y_{23}} v$
($\Delta \gg \sqrt{y_{14} y_{23}} v$)  
the mass splitting is dominated by electroweak
breaking (preserving) masses that we call the 
Linear (Quadratic) Case.
See the text for details.}
\label{fig:su4_spectrum_figure}
\end{figure}

An exact custodial $SU(2)$ symmetry implies the 
masses and interactions are 
symmetric with respect to the interchange $u \leftrightarrow d$.  
This means the Lagrangian parameters satisfy
\begin{eqnarray}
y_{14}^u = y_{14}^d \equiv y_{14}, & \qquad &
y_{23}^u = y_{23}^d \equiv y_{23}, \label{eq:custodiallimit} \\
M_{34}^u & = M_{34}^d & \equiv M_{34} \, . \nonumber
\end{eqnarray}
Defining the overall vector-like mass scale $M$ and difference 
$\Delta$ to be \footnote{We assume $\Delta < M$,
such that fermion masses remain positive, to avoid further fermion field
rephasings.}
\begin{eqnarray}
M      \equiv  \frac{M_{12} + M_{34}}{2} \qquad
\Delta \equiv  \left| \frac{M_{12} - M_{34}}{2} \right| \, ,
\end{eqnarray}
the dark fermion mass eigenvalues are
\begin{eqnarray}
M_{1,2} = M \mp \sqrt{\Delta^2 + \frac{y_{14} y_{23} v^2}{2}} \, .
\end{eqnarray}
No $u$ or $d$ labels are necessary, since custodial $SU(2)$ symmetry
implies that there is one pair of Dirac fermions with electric charge 
$Q = (+1/2,-1/2)$ with mass $M_1$ (the lightest pair), as well as a 
second pair of Dirac fermions with electric charge $Q = (+1/2,-1/2)$ 
with mass $M_2$ (the heavier pair).  The spectrum is illustrated
in Fig.~\ref{fig:su4_spectrum_figure}. 

In the limit $y_{14},y_{23} \rightarrow 0$, the fermions acquire purely 
vector-like masses,
and thus the chiral condensate of the dark force is aligned
to a purely electroweak-preserving direction.  
In order that the chiral condensate's
electroweak-preserving orientation is not significantly disrupted, 
we consider small electroweak breaking masses, $y_{14} v, y_{23} v \ll M$.

This leaves two distinct regimes for the spectrum, 
depending on the relative sizes 
of $\sqrt{y_{14} y_{23}} v$ and $\Delta$.

\subsection{Approximately symmetric mass matrices}

A second simplification, useful to analytically and numerically 
evaluate our results, is to take $y_{14} \simeq y_{23}$. The 
mass matrices Eqs.~(\ref{eq:upmassmatrix},\ref{eq:downmassmatrix}) 
are approximately symmetric.  Specifically, we can write
\begin{equation} \label{eq:approxsymmetric}
y_{14} = y + \eps_y \, , \quad y_{23} = y - \eps_y \, , \quad
|\eps_y| \ll |y| \, . 
\end{equation}
and expand in powers of $\epsilon_y$.  For example, 
the dark fermion masses become simply
\begin{eqnarray}
M_{1,2} = M \mp \sqrt{\Delta^2 + \frac{y^2 v^2}{2}} \, .
\end{eqnarray}
to leading order in $O(\epsilon_y)$.  

The distinct regimes are thus $y v \gg \Delta$ and $y v \ll \Delta$.
In the Linear Case $y v \gg \Delta$, 
electroweak symmetry breaking is (dominantly) responsible
for the mass \emph{splitting} between $\Psi^{u,d}_1$ and $\Psi^{u,d}_2$.  
In the Quadratic Case $y v \ll \Delta$, the splitting is dominantly 
attributed to the vector-like mass splitting $\Delta$.  As we shall see, 
the primary distinction between
these two cases is in the Higgs coupling
to the fermion mass eigenstates: proportional to $y$ for the Linear Case
and $y^2$ for the Quadratic Case, hence the case names.  
A similar observation was also found 
in Ref.~\cite{Hill:2014yka}. 

From this point forward unless noted otherwise, we assume the 
fermion mass parameters satisfy an exact custodial $SU(2)$ 
and the mass matrices are approximately symmetric.

\section{Light Non-Singlet Meson Phenomenology}
\label{sec:meson}

Theories with new fermions that transform under vector-like
representations of the electroweak group generically have
enlarged global flavor symmetries that can prevent decay of
the lightest non-singlet mesons and baryons.  In the case of dark baryons, 
this is a feature, providing the rationale for the stability
of the lightest dark baryon of the theory.

In the case of the lightest non-singlet mesons, this can be problematic,
since some of these mesons carry electric 
charge.\footnote{We use the term ``lightest mesons'' and not 
``pions'' since the would-be global symmetry that protects 
pion masses is completely broken by the dark fermion vector-like 
masses.  Nevertheless, we use the symbol $\Pi$ to denote the corresponding fields.}
Stable integer charged mesons are strongly constrained from collider searches as well as cosmology.
One solution 
is to postulate additional higher dimensional operators that
connect a dark fermion pair with a Standard Model 
fermion pair \cite{Kilic:2009mi,Buckley:2012ky}. 
This must be carefully done to avoid also
writing operators that violate the approximate global 
symmetries protecting the stability of the dark matter. 
In the stealth dark matter model, however, electroweak symmetry
breaking can provide the source of global flavor symmetry breaking,
leading to the decay of the lightest charged mesons.  (We will not discuss the lightest neutral mesons, but they are generically more difficult to produce in colliders, and they will decay through essentially the same mechanism as we describe for the charged mesons.)

The lightest electrically charged mesons are composed 
dominantly of the dark fermion pairs 
$\Pi^+ = (\overline{\Psi^d_1} \Psi_1^u)$ and 
$\Pi^- = (\overline{\Psi^u_1} \Psi_1^d)$. 
We can estimate the lightest meson lifetime by generalizing 
pion decay of QCD to our model.  The relevant matrix element is
(see, e.g., \cite{Donoghue:1992dd})
\begin{eqnarray}
\langle 0 | j^\mu_{\pm,{\rm axial}} | \Pi^{\pm} \rangle &=& i f_\Pi p^\mu \, ,
\end{eqnarray}
where $f_\Pi$ is the ``pion decay constant'' associated with
the dark force in this paper.  
The axial part of the electroweak current can be read off 
from the electroweak currents given in 
Eqs.~(\ref{eq:jplus}),(\ref{eq:jminus})
\begin{eqnarray}
j^\mu_{+,{\rm axial}} &\supset& 
  c_{\rm axial} 
  \overline{\Psi^u_1} \gamma^\mu \gamma_5 \Psi^d_1
\end{eqnarray}
where 
\begin{eqnarray}
c_{\rm axial} &=& \frac{c_1^u c_1^d - c_2^u c_2^d}{\sqrt{2}} 
\end{eqnarray}
and $j^\mu_{-,{\rm axial}}$ is identical upon $u \leftrightarrow d$. 
In the custodial limit, Eq.~(\ref{eq:custodiallimit}), 
the axial coefficient is
\begin{equation}
c_{\rm axial} = \frac{(y_{14}^2 - y_{23}^2) v^2}{\sqrt{2 
(8 M^2 + (y_{14} - y_{23})^2 v^2) (8 \Delta^2 + (y_{14} + y_{23})^2 v^2)}} \, .
\end{equation}
Some insight can be gained using approximately symmetric
mass matrices, Eq.~(\ref{eq:approxsymmetric}).
We then obtain
\begin{eqnarray}
c_{\rm axial} &=& \frac{\eps_y y v^2}{2 M \sqrt{2 \Delta^2 + y^2 v^2}} 
\\ &\simeq& \frac{\eps_y v}{2 M} \times \left\{ 
  \begin{array}{ll}
  1 & \quad \mbox{Linear Case} \\
  y v/(\sqrt{2} \Delta) & \quad \mbox{Quadratic Case.} 
  \end{array}
\right. \nonumber
\end{eqnarray}
The decay width can be obtained from pion decay of QCD by
replacing $V_{ud}$ in the Standard Model with $c_{\rm axial}$ 
for the dark mesons. 
Since the 
charged dark mesons of this model are much heavier than 
the QCD pions, there are many possible decay modes.
For a general decay 
into a Standard Model doublet $(f \, f')$,
assuming $m_f \gg m_{f'}$, the decay width is
\begin{eqnarray}
\Gamma(\Pi^+ \ra f \overline{f}') &=&
\frac{G_F^2}{4 \pi} f_\Pi^2 m_f^2 m_{\Pi} c_{\rm axial}^2
\left( 1 - \frac{m_f^2}{m_{\Pi}^2} \right) \, .
\end{eqnarray}
If $m_{\Pi} > m_t + m_b$, the dominant decay mode is expected
to be $\Pi^+ \ra t\overline{b}$, 
otherwise $\Pi^+ \ra \tau^+ \nu_\tau$ and $\Pi^+ \ra \bar{s} c$, with branching ratios of roughly 70\% and 30\% respectively.
Note that the decay width has several enhancement factors
relative to the QCD pion decay width
\begin{equation}
\frac{\Gamma(\Pi^+ \ra f \overline{f}')}{\Gamma(\pi \ra \mu^+ \nu_\mu)}
\simeq \frac{c_{\rm axial}^2}{|V_{ud}|^2} \left( \frac{f_\Pi}{f_\pi} \right)^2
\left( \frac{m_f}{m_\mu} \right)^2
\left( \frac{m_\Pi}{m_\pi} \right)
\end{equation}
where for simplicity we have neglected kinematic suppression. 
As an example, if $f_\Pi \simeq m_\Pi \simeq v$, we find the 
lightest charged dark mesons decay faster than QCD charged pions 
so long as $c_{\rm axial} \gsim 10^{-8}$.  This is easy to satisfy 
with small Yukawa couplings and dark fermion masses at or 
beyond the electroweak scale.  

We can now make some comments about existing collider constraints
on non-singlet mesons.  The lightest charged mesons 
$\Pi^{\pm}$ can be pair produced in particle colliders 
through the Drell-Yan process, and will decay through annihilation 
of the constituent fermions into a $W$ boson.  Because the Drell-Yan production is mediated by a photon and the mesons have unit electric charge, the production cross-section is substantial, leading to robust bounds from LEP-II.  For charged states near the LEP-II energy
threshold, the dominant decay mode is expected to be $\Pi^+ \ra \tau^+ \nu_\tau$ as noted above.
Reinterpreting the LEP-II bound from the pair production of 
supersymmetric partners to the tau (with the stau decaying
into a tau and a nearly massless gravitino), we find 
$m_{\Pi} \gsim 86.6$~GeV
\cite{Heister:2001nk,Heister:2003zk,Abdallah:2003xe,Achard:2003ge,Abbiendi:2004gf}. 
Stronger bounds from the LHC may be possible, although existing searches do not yet give any significant constraints on the charged mesons \cite{Buckley:2012ky}; 
we briefly highlight the signals in the discussion.

Using our lattice results from Ref.~\cite{Appelquist:2014jch},
we can translate the experimental bound on the mass of the pseudoscalar 
meson into a bound on the baryon mass, 
$m_B > 245, 265, 320$~GeV when the ratio of the pseudoscalar mass
to the vector meson mass is $m_{\Pi}/m_V = 0.77, 0.70, 0.55$. 

\section{Contributions to Electroweak Precision Observables}
\label{sec:pew}

Stealth dark matter contains dark fermions that acquire 
electroweak symmetry breaking contributions to their masses.
Consequently, there are contributions to the electroweak precision
observables of the Standard Model, generally characterized by
$S$ and $T$ \cite{Peskin:1990zt,Peskin:1991sw}. 
In the custodial $SU(2)$ limit, Eq.~(\ref{eq:custodiallimit}), 
the contribution to $T$ vanishes.
There is a contribution to $S$, controllable
through the relative size of the electroweak breaking and
electroweak preserving masses of the dark fermions.

The $S$ parameter is defined in terms of momentum derivatives of
current-current correlators \cite{Peskin:1990zt,Peskin:1991sw},
\begin{eqnarray}
S &\equiv& 16\pi \Pi^\prime_{3Y}(0) \\
  &=&      \frac{d}{dq^2} \left[ \frac{16\pi}{3}
                            \left(g^{\mu\nu}-\frac{q^\mu q^\nu}{q^2} \right)
                            X^{\mu\nu}(q^2)\right]_{q^2=0} \nonumber \\
X^{\mu\nu}(q^2) 
  &\equiv& \int d^4x\  e^{-i q x} \langle j^\mu_3(x) j^\nu_Y(0) \rangle,
\end{eqnarray}
where the currents $ j^\mu_3(x)$ and $ j^\nu_Y(x)$ for the stealth dark 
matter model are defined in Eqs.~(\ref{eq:j3}) and (\ref{eq:jY}).  
After some algebra and
identifications of symmetric contractions, these definitions of the 
currents in terms of 4-component fermion fields lead to the 
current-current correlator.  In the custodial limit, we obtain
\begin{eqnarray}
2\langle & & j^\mu_3(x) j^\nu_Y(0) \rangle 
  = c_1^2 s_1^2 \left(   {}^{11}G^{\mu\nu}_{LL}
                         + {}^{22}G^{\mu\nu}_{LL}
                         - {}^{12}G^{\mu\nu}_{LL}
                         - {}^{21}G^{\mu\nu}_{LL} \right) \nonumber \\
& &{} + c_2^2 s_2^2 \left(   {}^{11}G^{\mu\nu}_{RR}
                            + {}^{22}G^{\mu\nu}_{RR}
                            - {}^{12}G^{\mu\nu}_{RR}
                            - {}^{21}G^{\mu\nu}_{RR} \right) \nonumber \\
& &{} + c_1^2s_2^2 \left(   {}^{11}G^{\mu\nu}_{LR}
                          + {}^{22}G^{\mu\nu}_{RL} \right) 
      + c_2^2s_1^2 \left(   {}^{11}G^{\mu\nu}_{RL}
                          + {}^{22}G^{\mu\nu}_{LR} \right) \nonumber \\
& &{} - c_1c_2s_1s_2 \left(   {}^{12}G^{\mu\nu}_{LR}
                             + {}^{12}G^{\mu\nu}_{RL}
                             + {}^{21}G^{\mu\nu}_{LR}
                             + {}^{21}G^{\mu\nu}_{RL} \right),
\end{eqnarray}
where the connected contributions to the correlation functions are given by
\begin{equation}
{}^{ij} G^{\mu\nu}_{AB} 
  \equiv \left. \langle \bar{\Psi}^u_i \gamma^\mu P_A 
                  \Psi^u_j \bar{\Psi}^u_j \gamma^\nu P_B \Psi^u_i     
          \rangle \right|_{\text{connected}} \, . \label{eq:curr_corr}
\end{equation}
Here, $A,B = L,R$ and the flavor indices $i,j = 1,2$, where it is 
understood that the flavors labeled $2$  have larger fermion masses 
than the flavors labeled $1$.  Since the $u,d$ flavors have the same mass, 
the $u$ and $d$ labels are interchangeable (i.e. everything is 
written in terms of the $u$ flavors).  

We can obtain expressions for the mixing angle coefficients. 
Like the case of light meson decay, if we consider an
approximately symmetric mass matrix, with Yukawa couplings
given by Eq.~(\ref{eq:approxsymmetric}), all of the 
mixing angle coefficients are approximately equal to each other, 
differing only at first order in $\epsilon_y$, i.e., 
\begin{eqnarray}
c_1^2 s_1^2 &\simeq& c_2^2 s_2^2 \simeq 
                     c_1^2 s_2^2 \simeq c_2^2 s_1^2 \simeq c_1 c_2 s_1 s_2 
                \nonumber \\
            &=& \frac{1}{4} \frac{y^2 v^2}{y^2 v^2 + 2 \Delta^2} 
                \left[ 1 + O(\epsilon_y) \ldots \right] \nonumber \\
            &\simeq& \frac{1}{4} \times \left\{ 
  \begin{array}{ll}
  1 & \quad \mbox{Linear Case} \\
  y^2 v^2/(2 \Delta^2) & \quad \mbox{Quadratic Case.} 
  \end{array}
\right. 
\end{eqnarray}
In the Linear Case, the mixing angles are approximately equal  
$c_1 \simeq s_1 \simeq c_2 \simeq s_2 \simeq 1/\sqrt{2}$.
In the Quadratic Case, all of the contributions to the $S$ parameter 
are suppressed by $(y v/\Delta)^2$.  To calculate the $S$ parameter
in general requires lattice methods, paying close attention to the 
heavy-light splitting of the fermions, $M_2 - M_1$. 
To a first approximation we expect that in the limit of small mass
splitting, $M_2 - M_1 \ll M$,
\begin{equation}
   G_{AB}^{\mu\nu} \equiv {}^{11}G_{AB}^{\mu\nu} \simeq
{}^{22}G_{AB}^{\mu\nu} \simeq {}^{12}G_{AB}^{\mu\nu} \simeq
{}^{21}G_{AB}^{\mu\nu}.
\end{equation}
This gives for the current--current correlator
\begin{align}
   2 \langle j_3^{\mu}(x) j_Y^{\nu}(0)\rangle & \simeq \left[c_1^2 s_2^2
+ c_2^2 s_1^2 - 2c_1 c_2 s_1 s_2\right] G_{LR}^{\mu\nu} \nonumber \\
   & \simeq \frac{\epsilon_y^2 v^2}{2M^2} G_{LR}^{\mu\nu},
\end{align}
where all of the $G_{LL}$ and $G_{RR}$ contributions self-cancel.
Hence, we see that the contribution to the $S$ parameter
is suppressed as $M \gg v$ or $\epsilon_y \ll 1$,
as expected.

\section{Fermion Couplings to the Higgs Boson}
\label{sec:fermionhiggs}

In terms of the gauge-eigenstate fields, the interactions of the Higgs boson 
with the dark-sector fermions are, in matrix notation,
\begin{eqnarray} 
\mathcal{L} &\supset& - \frac{h}{\sqrt{2}} (F_1^u \;\; F_3^u) 
                        \left( \begin{array}{cc}
                        0 & y_{14}^u \\ y_{23}^u & 0
                        \end{array} \right)
                        \spinvec{F_2^d}{F_4^d} \nonumber \\
& &{}  + \frac{h}{\sqrt{2}} (F_1^d \;\; F_3^d) 
                        \left( \begin{array}{cc}
                        0 & y_{14}^d \\ y_{23}^d & 0
                        \end{array} \right)
                        \spinvec{F_2^u}{F_4^u} \nonumber \\
    & &{} + h.c. \, .
\end{eqnarray}
These matrices are not simultaneously diagonalizable with the mass matrices,
Eqs.~(\ref{eq:upmassmatrix}),(\ref{eq:downmassmatrix}).
This means that the Higgs boson in 
general has off-diagonal, ``dark flavor-changing'' interactions with the 
mass eigenstate fields.  Explicitly, we find in terms 
of the mixing angles
\begin{widetext}
\begin{eqnarray}
\mathcal{L} \supset 
    \frac{h}{\sqrt{2}}
    \left( \begin{array}{cc} \overline{\Psi}_1^u & \overline{\Psi}_2^u 
           \end{array} \right) 
    \left( \begin{array}{cc}
    c_1^u s_2^u \, y_{14}^u + s_1^u c_2^u \, y_{23}^u 
  &~~ c_1^u c_2^u \, y_{14}^u - s_1^u s_2^u \, y_{23}^u \\
    c_1^u c_2^u \, y_{23}^u - s_1^u s_2^u \, y_{14}^u 
  &~~ - s_1^u c_2^u \, y_{14}^u - c_1^u s_2^u \, y_{23}^u 
    \end{array} \right) 
    \spinvec{\Psi_1^u}{\Psi_2^u}  
     + (u \leftrightarrow d) \, .
\end{eqnarray}
In the custodial $SU(2)$ limit, we can drop the $u$ and $d$ labels
since the Higgs coupling matrix is identical for both sets of fields.
If we further take the limit of an approximately symmetric 
mass matrix, Eq.~(\ref{eq:approxsymmetric}), the Higgs couplings
simplify to
\begin{equation}
\mathcal{L} \supset 
    \frac{y h}{M_2 - M_1}
    \left( \begin{array}{cc} \overline{\Psi}_1 & \overline{\Psi}_2 
           \end{array} \right) 
    \left[ \left( \begin{array}{cc}
           y v & -\sqrt{2} \Delta \\
           -\sqrt{2} \Delta & - y v
    \end{array} \right) + O(\epsilon_y) \right] 
    \spinvec{\Psi_1}{\Psi_2} \, . 
\end{equation}
\end{widetext}
We observe both diagonal and off-diagonal Higgs couplings to
the fermions.  The off-diagonal dark flavor-changing interactions 
vanish in the limit 
$\Delta \rightarrow 0$ and $\epsilon_y \rightarrow 0$. 
In this limit an enhanced flavor symmetry among the fermions is restored, 
and the analogue of the GIM mechanism forbids such interactions at tree-level.
The off-diagonal Higgs couplings lead to an inelastic scattering 
cross section when a single Higgs is exchanged.  This is highly
suppressed unless the mass difference $M_2 - M_1$ is near the
(non-relativistic) kinetic energy of the dark matter in galaxy.
Two off-diagonal Higgs couplings can be combined in a loop
involving one heavier dark fermion and double Higgs exchange,
but this is suppressed by the square of the Higgs couplings
times a loop factor, as well as by the mass of the heavier fermions. 

The single Higgs coupling to the lightest fermions is finally
\begin{eqnarray}
\mathcal{L} &\supset& y_\Psi h \overline{\Psi}_1 \Psi_1 
\end{eqnarray}
where
\begin{eqnarray}
y_\Psi &=& \frac{y^2 v}{M_2 - M_1} + O(\epsilon_y) \nonumber \\
       &\simeq& \begin{dcases*}
\frac{y}{\sqrt{2}} & \mbox{Linear Case} \\
\frac{y^2 v}{2 \Delta} & \mbox{Quadratic Case.} 
\end{dcases*}
\end{eqnarray}
(Note also that the single Higgs coupling to the heaviest fermions 
$\Psi_2$ is identical up to an overall sign.)
Depending on the relative size of $y v$ and $\Delta$, the Higgs boson
couples linearly or quadratically proportional to the 
Yukawa coupling $y$.  The additional suppression of
$y v/\Delta$ in the Quadratic Case will imply 
that spin independent scattering through single Higgs exchange
can be significantly weaker when the mass difference between
the lightest and heaviest fermions is dominated by the
electroweak preserving mass $\Delta$.

\section{Direct detection bounds from Higgs exchange}
\label{sec:boundshiggs}

In a previous paper \cite{Appelquist:2014jch}, we determined the
model-independent bounds on direct detection from Higgs exchange
for a scalar baryon of $SU(4)$.  The model-independent result
was expressed in terms of the effective Higgs coupling to the baryon
\begin{equation}
g_B = \frac{m_B}{v} \alpha f^{(B)}_{f} \, . \label{gB-eq} 
\end{equation}
The first factor, the baryon mass $m_B$ (divided by the electroweak VEV), 
as well as the third factor
\begin{equation}
f_f^{(B)} = \frac{\langle B| M_1 \overline{\Psi}_1 \Psi_1 | B \rangle}{m_B}
          = \frac{M_1}{m_B} \frac{\partial m_B}{\partial M_1}
\nonumber 
\end{equation}
are extracted from our lattice results \cite{Appelquist:2014jch}. 
The second factor
\begin{equation}
\alpha \equiv \frac{v}{M_1} 
              \frac{\partial \, M_1(h)}{\partial \, h}\bigg|_{h=v} 
      \simeq \begin{dcases*} 
               \frac{y v}{\sqrt{2} M_1} & \text{Linear Case} \\
               \frac{(yv)^2}{2 M_1 \Delta} & \text{Quadratic Case}
             \end{dcases*}
\label{eq:alpha}
\end{equation}
provides the effective coupling of the Higgs boson to the fermions
(multiplied by $v/M_1$), and we have evaluated the derivative
for the two cases in our model.

\begin{figure}[t] 
\includegraphics[width=0.45\textwidth]{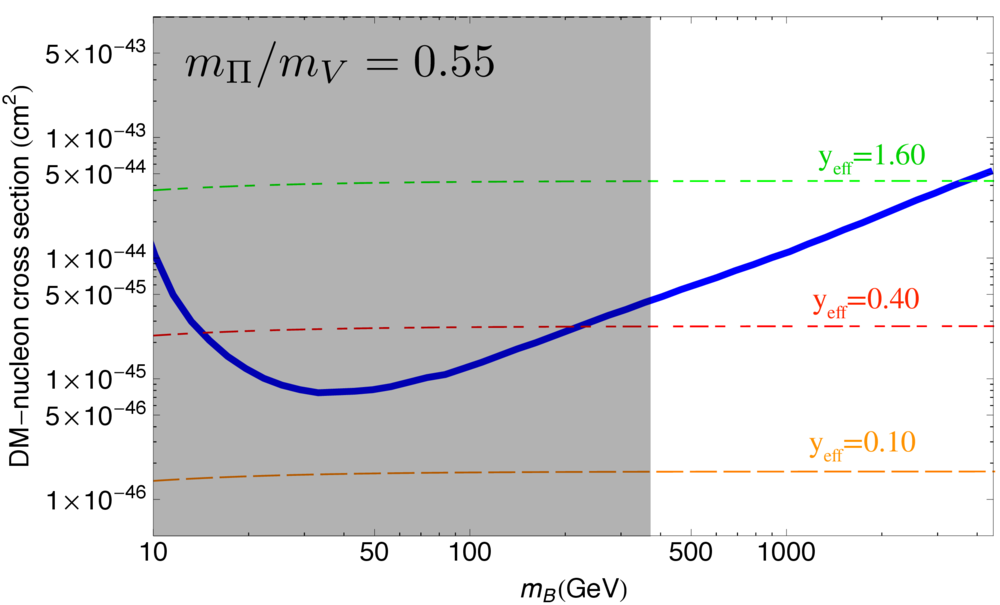}
\includegraphics[width=0.45\textwidth]{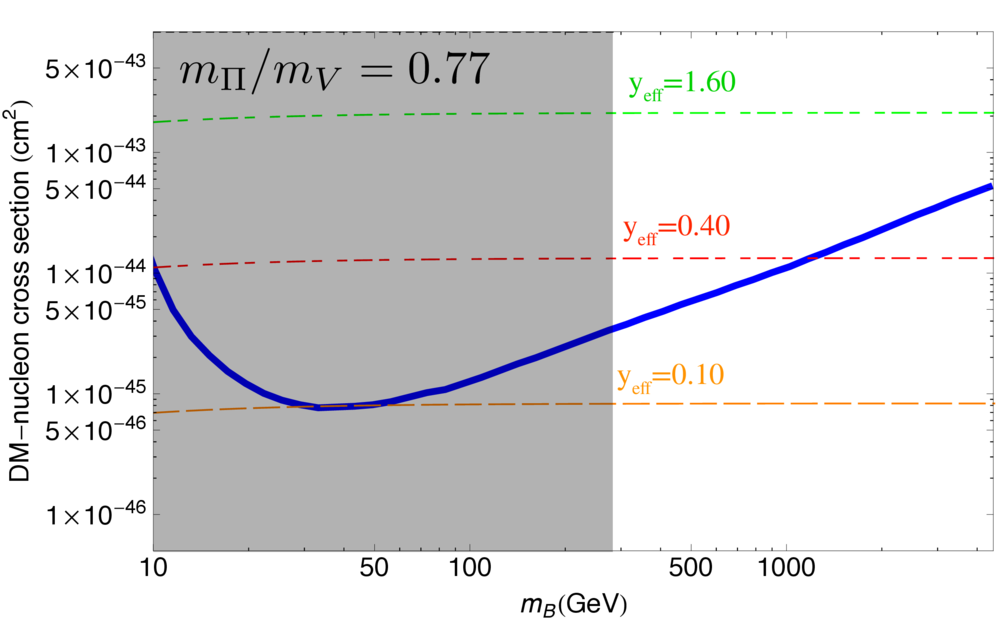}
\includegraphics[width=0.45\textwidth]{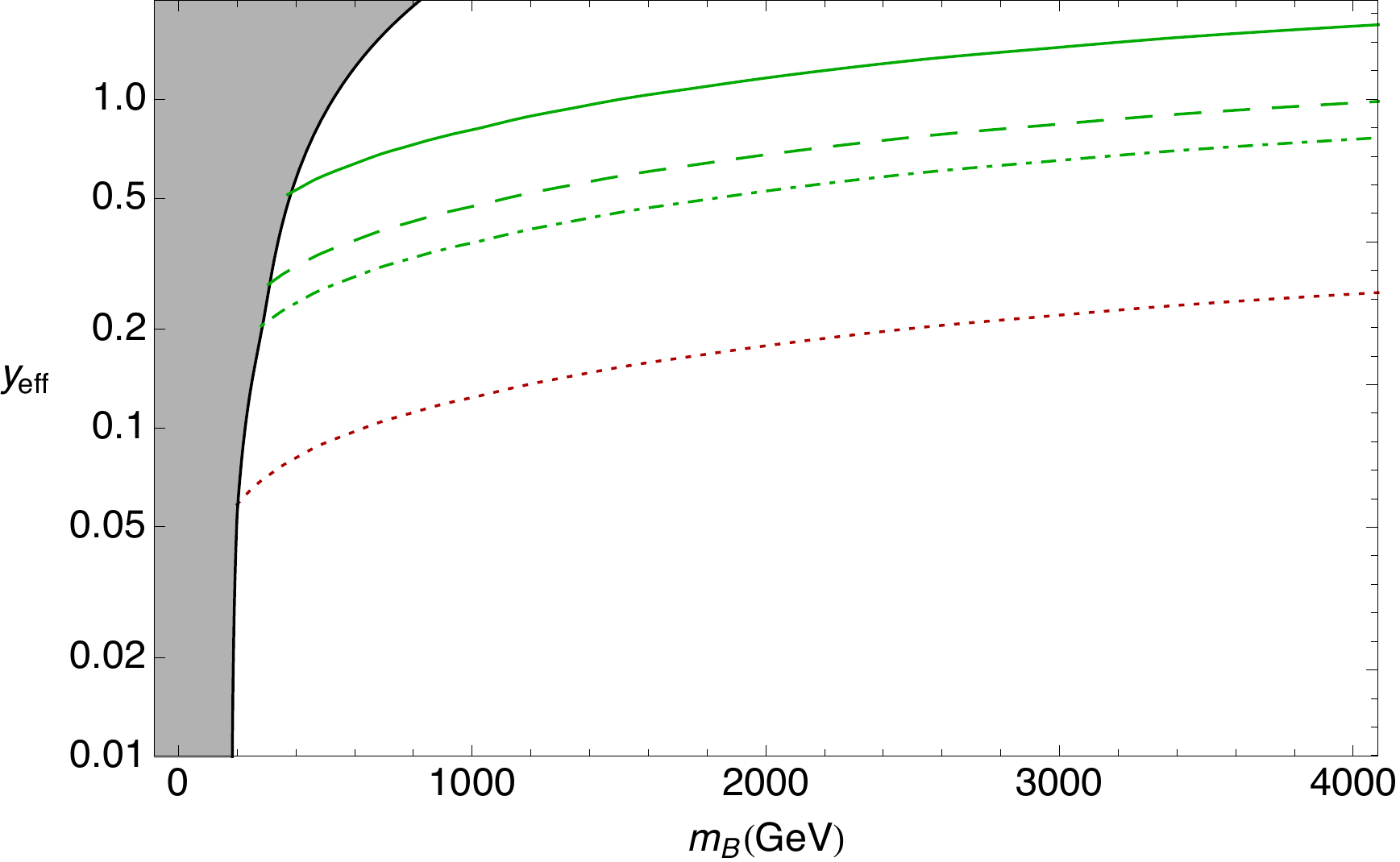}
\caption{Constraints on the stealth dark matter model in the Linear Case 
of the model. The top and middle figures show the predicted values for the 
smallest and largest fermion mass explored in our simulations 
(corresponding to the pseudoscalar to vector mass ratio 
$m_{\Pi}/m_V = 0.55, 0.77$) as well as LUX bounds.  
Various $y_{\text{eff}}$ values are plotted on the figure, 
where $y_{\text{eff}} \approx y m_B/M_1$ in this case.  
The dark grey region is excluded by the 
LEP constraints on charged dark mesons.
The bottom figure displays the 
maximum $y_{\text{eff}}$ allowed for a given dark matter mass.  
Each of the green curves represents a different fermion mass in 
the lattice calculation, $m_{\Pi} / m_V = 0.55, 0.7, 0.77$ from top to bottom, and the bottom red curve is the result in the heavy fermion limit.} 
\label{fig:lin}
\end{figure}

\begin{figure}[t]
\includegraphics[width=0.45\textwidth]{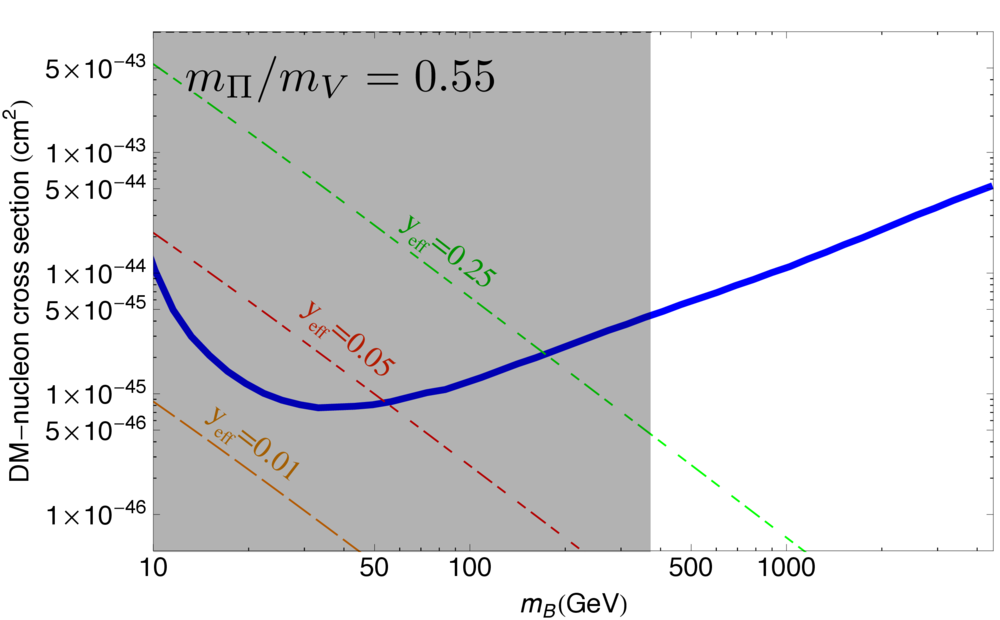}
\includegraphics[width=0.45\textwidth]{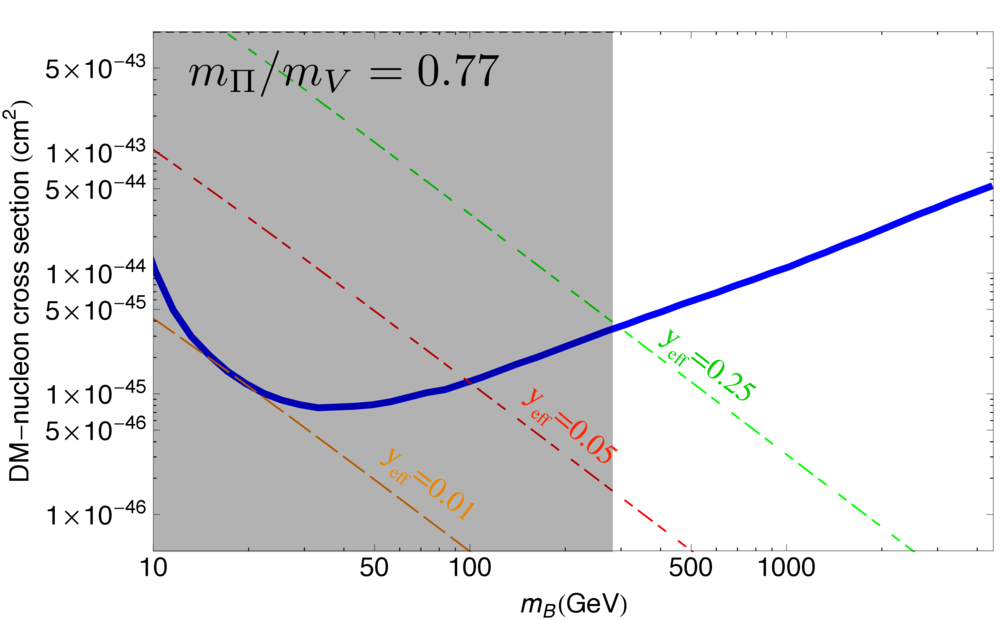}
\includegraphics[width=0.45\textwidth]{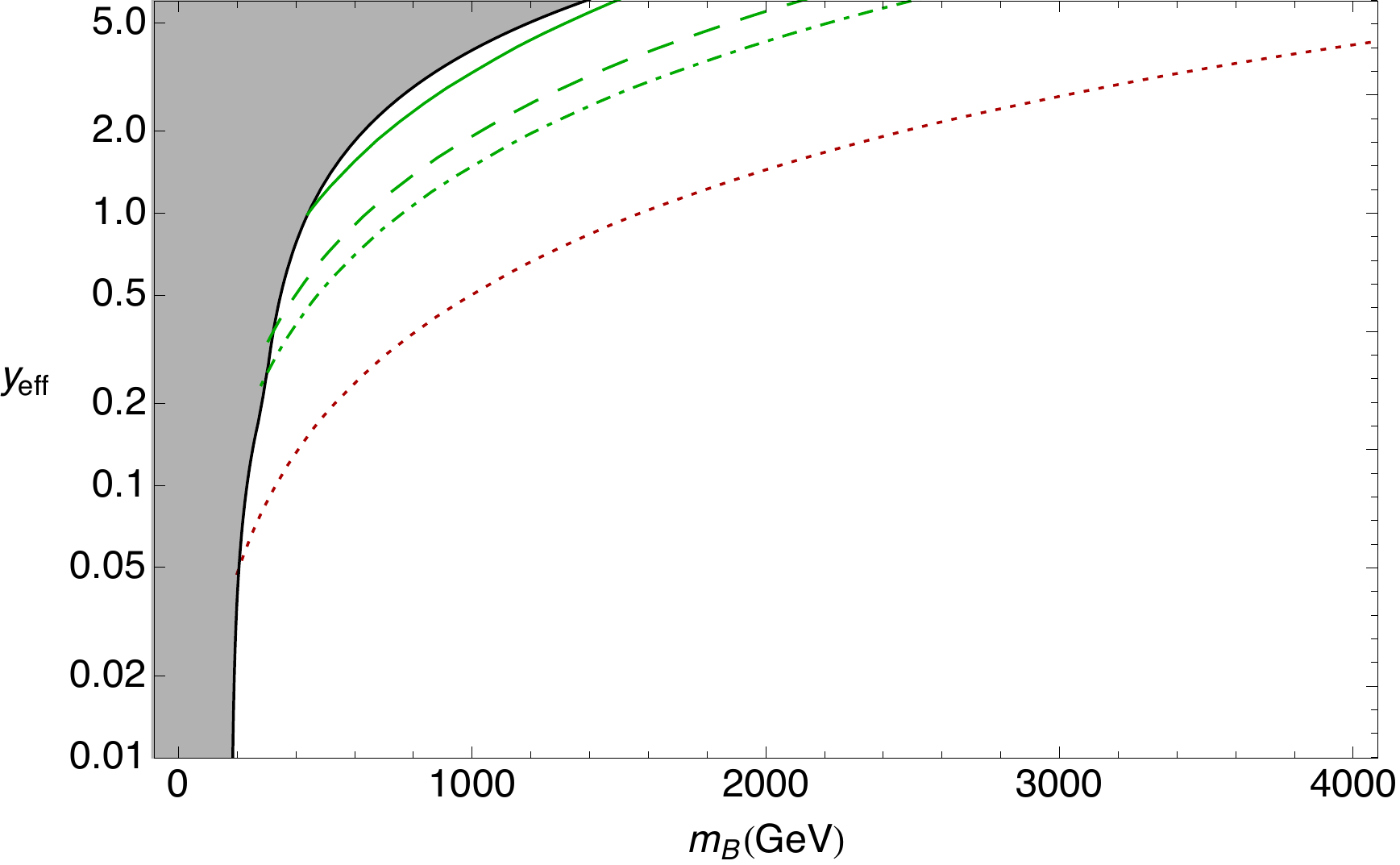}
\caption{Same as Fig.~\ref{fig:lin} but for the Quadratic Case of 
the model.  In this case, $y_{\text{eff}} \approx  y m_B/\sqrt{M_1 \Delta}$.}
\label{fig:quad}
\end{figure}

Unfortunately, we cannot directly apply our previous results
on constraints in $\alpha$-$m_B$ space to the parameters
of the stealth dark matter model.  This is because we do not know 
the dark fermion mass, $M_1$, independent of the lattice regularization
scheme.  We can, however, construct a regularization-independent
parameter, the effective Yukawa coupling $y_{\rm eff}$, 
that is closely related to the model parameters:
\begin{equation}
y_{\text{eff}} \equiv 
  \begin{dcases*} 
    y \frac{m_B}{\sqrt{2} M_1}        & \text{Linear Case} \\ 
    y \frac{m_B}{\sqrt{2 \Delta M_1}} & \text{Quadratic Case.}  
  \end{dcases*}
\label{eq:yeff}
\end{equation}
The $\alpha$ parameter is therefore
\begin{equation}
\alpha \simeq 
  \begin{dcases*} 
    y_{\rm eff}   \frac{v}{m_B}     & \text{Linear Case} \\
    y_{\rm eff}^2 \frac{v^2}{m_B^2} & \text{Quadratic Case.} 
  \end{dcases*}
\label{eq:alpha2}
\end{equation}
Recasting our previous constraints in $\alpha$-$m_B$ space
into $y_{\rm eff}$-$m_B$ space, we can identify the region
of parameter space that remains viable.  
The constraints for the Linear Case are shown in 
Fig.~\ref{fig:lin} and the Quadratic Case in Fig.~\ref{fig:quad}. 
In the top two plots for the respective figures, 
the region above the LUX bounds represents the excluded 
parameter space for the model at a given dark matter mass ($m_B$) 
and effective Yukawa coupling ($y_{\text{eff}}$).  
The figures show a clear qualitative trend in how the 
predictions change as a function of dark matter mass.  
In particular, the cross-section is independent of $m_B$ 
for the Linear Case and inversely proportional to $m_B$
in the Quadratic Case.  
The bottom plots in Figs.~\ref{fig:lin},\ref{fig:quad} shows 
the maximum $y_{\rm eff}$ allowed for a given dark matter mass. 
By increasing the splitting 
$\Delta$ between the vector-like mass terms, significantly more
$y_{\text{eff}}$ parameter space becomes available.


\section{Abundance}
\label{sec:abundance}

We now provide a brief discussion of the relic abundance of
stealth dark matter.  In the regime where the dark fermions
have masses comparable to the confinement scale of the dark force,
calculating the relic abundance is an intrinsically 
strongly-coupled calculation.  Unfortunately, this calculational
difficulty is not easily overcome with lattice simulations, 
due to the different initial and final states.  
Nevertheless, it is straightforward to see that the 
relic abundance \emph{can} match the cosmological abundance through 
at least two distinct mechanisms that lead to two different mass scales 
for stealth dark matter.  In this section we discuss obtaining
the abundance of stealth dark matter through thermal freezeout,
leading to a symmetric abundance of dark baryons and anti-baryons. 
Separately, we consider the possibility of an asymmetric abundance
generated through electroweak sphalerons.
 
\subsection{Symmetric Abundance}

In the early universe at temperatures well above the confinement scale 
of the $SU(4)$ dark gauge force, the dark fermions are in thermal
equilibrium with the thermal bath through their electroweak
interactions.  As the universe cools to temperatures
below the confinement scale, the degrees of freedom
change from dark fermions and gluons into 
the dark baryons and mesons of the low energy description.
Some of the dark mesons carry electric charge, and so 
the dark mesons remain in thermal equilibrium with the 
Standard Model quarks, leptons, and gauge fields.  
Since the dark baryons are strongly coupled to the dark mesons,
they also are kept in thermal equilibrium.
As the temperature of the universe falls well below the
mass of the dark baryons, they annihilate into dark mesons
that subsequently thermalize and decay (or decay then thermalize)
into Standard Model particles.  The symmetric abundance of 
dark baryons is therefore determined by the annihilation rate of 
dark baryons into dark mesons.  

The annihilation of dark baryons to dark mesons is
a strongly coupled process.
We expect $B^* B \rightarrow \Pi \, \Pi$, 
$B^* B \rightarrow 3 \, \Pi$, 
and $B^* B \rightarrow 4 \, \Pi$, 
(and to possibly more mesons if kinematically allowed) 
to occur, but we do not know the dominant annihilation channel.
If the 2-to-2 process $B^* B \rightarrow \Pi \, \Pi$ dominates, 
one approach is to use partial wave unitarity to estimate the 
thermally averaged annihilation rate \cite{Griest:1989wd,Blum:2014dca},
\begin{eqnarray}
\langle \sigma v \rangle 
  &\sim& \frac{4 \pi \langle v^{-1} \rangle}{m_B^2} \, ,
\end{eqnarray}
where $\langle v^{-1} \rangle \simeq 2.5$ at 
freezeout \cite{Blum:2014dca}.   Matching this cross section to 
the required thermal relic abundance yields
$m_B \sim 100$~TeV\@.
An alternative approach is to use naive dimensional analysis 
\cite{Manohar:1983md,Luty:1997fk,Cohen:1997rt}, which appears to 
lead to a larger dark matter mass. 

If the 2-to-3 or 2-to-4 processes dominate instead, 
the additional phase space and kinematic suppression
lowers the annihilation rate, and therefore lowers the 
scalar baryon mass needed to obtain the cosmological abundance.  
For recent work that has considered the thermal relic abundance 
in multibody processes, see \cite{Hochberg:2014dra,Hochberg:2014kqa}. 
Suffice it to say a symmetric thermal abundance 
of dark baryons will match the cosmological abundance
for a relatively large baryon mass that is of order 
tens to hundreds of TeV\@.

\subsection{Asymmetric Abundance}

Early work on technibaryons demonstrated that strongly-coupled 
dark matter could arise from an asymmetric abundance 
\cite{Nussinov:1985xr,Chivukula:1989qb,Barr:1990ca,Barr:1991qn,Kaplan:1991ah}.
The main ingredient to obtain the correct cosmological abundance
involved the electroweak sphaleron -- the non-perturbative solution 
at finite temperature that allows for transitions between vacua with 
different\footnote{In this section, $B$ refers to baryon number and is to not be confused with the field defined earlier} $B+L$ numbers.\footnote{In
addition, an asymmetric abundance could be generated through other 
mechanisms, see Ref.~\cite{Kaplan:2009ag}, in which case 
the mass scales and parameters depend on the details of 
the particular mechanism.}  In the early universe, at temperatures
much larger than the electroweak scale, electroweak sphalerons
are expected to violate one accidental global symmetry, 
$B+L+D$ number, leaving $B-L$ and $B-D$ numbers unaffected
\cite{Barr:1991qn,Kaplan:1991ah,Kribs:2009fy}.  
Here $D$ number is proportional to the dark baryon number, 
with some appropriate normalization (for examples, see 
\cite{Kaplan:1991ah,Kribs:2009fy}). 

Given a baryogenesis mechanism,
the electroweak sphalerons redistribute baryon number into
lepton number and dark baryon number.  As the universe cools,
the mass of the technibaryon becomes larger than the temperature
of the Universe.  Eventually, the universe cools to the point
where electroweak sphalerons ``freeze out'' and can no longer 
continue exchanging $B$, $D$, and $L$ numbers.
The residual abundance of dark baryons is 
$\rho \sim m_B n_B$ where the number density is 
proportional to $\exp[-m_{\rm B}/T_{\rm sph}]$, where $T_{\rm sph}$ is the temperature
at which sphaleron interactions shut off. 

If the baryon and dark baryon number densities are comparable,
the would-be overabundance of dark matter (from $m_B \gg m_{\rm nucleon}$)
is compensated by the Boltzmann suppression.  Very roughly,
$m_B \sim 1$-$2$~TeV
is the natural mass scale that matches the cosmological abundance
of dark matter \cite{Barr:1990ca}. 
A crucial component of the early 
technibaryon papers \cite{Nussinov:1985xr,Chivukula:1989qb,Barr:1990ca}
is that the technifermions were in a purely chiral representation 
of the electroweak group, like the fermions of the Standard Model.

In stealth dark matter, given an early  
baryogenesis mechanism 
(or other analogous mechanism to generate an asymmetry in a 
globally conserved quantity 
\cite{Barr:1991qn,Shelton:2010ta,Davoudiasl:2010am,Haba:2010bm,Buckley:2010ui,McDonald:2011zza,Blennow:2010qp,Falkowski:2011xh}), 
it is possible that electroweak sphalerons 
could also lead to the correct relic abundance of dark baryons 
consistent with cosmology.

There is one critical difference from the early technicolor models
(as well as the quirky dark matter model):  
The dark fermions in stealth dark matter have both
vector-like and electroweak symmetry breaking masses.
This leads to a suppression of the effectiveness
of the electroweak sphalerons by a factor of $\alpha$, 
c.f.~Eq.~(\ref{eq:alpha}), leading to a somewhat \emph{smaller} 
stealth baryon mass to obtain the correct relic abundance compared 
with a technicolor model (all other parameters equal).
A more quantitative estimate is complicated by several factors:
\begin{itemize}  
\item Determining how the electroweak sphaleron redistributes the 
conserved global charges in the presence of fermions that acquire 
both electroweak preserving and electroweak breaking masses.
To the best of our knowledge, this calculation has never been done.
\item Determining the precise temperature at which electroweak sphalerons 
shut off, in the presence of both the Standard Model and stealth 
dark matter degrees of freedom contributing to the thermal bath. 
\item The baryogenesis mechanism itself, that determines the 
initial $B-L$ and $B-D$ numbers.  
\end{itemize}

Given the exponential suppression of the asymmetric abundance
as the dark baryon mass is increased, it is clear that the
upper bound on the dark baryon mass is nearly the same as the
technibaryon calculation (updated to the current cosmological
parameters), when stealth dark fermions have vector-like masses
comparable to electroweak symmetry breaking masses.
(This case is, however, constrained by the $S$ parameter, 
see Sec.~\ref{sec:pew}).  We can therefore anticipate that
a range of stealth dark matter masses will be viable, up to about
a TeV.  More precise predictions require further detailed investigation
that is beyond the scope of this paper.

\section{Discussion}
\label{sec:discussion}

We have presented a concrete model, ``stealth dark matter",
that is a composite baryonic scalar of a new $SU(N_D)$ 
strongly-coupled confining gauge theory with dark fermions 
transforming under the electroweak group.  
Though the stealth dark matter model has a 
wide parameter space, we focused on dark fermion masses that
respect an exact custodial $SU(2)$.
Custodial $SU(2)$ implies the lightest 
bosonic baryonic composite is an electrically neutral scalar 
(and not a vector or spin-2) of the $SU(N_D)$ dark spectrum, 
and in addition does not have a charge radius.  
This yields an exceptionally ``stealthy'' dark matter candidate,
with spin-independent direct detection scattering 
proceeding only through Higgs exchange (studied in this paper)
and the polarizability interaction (studied in our companion paper 
\cite{Appelquist:2015zfa}).
Custodial $SU(2)$ also allows for stealth dark matter to 
completely avoid the constraints from the $T$ parameter.  
While contributions to the $S$ parameter are present, they 
are  suppressed by the ratio of
the electroweak symmetry breaking mass-squared divided by a 
vector-like mass squared of the dark fermions.
We also verified the lightest non-singlet mesons decay rapidly
(so long as $\epsilon_y \not= 0$), avoiding any cosmological 
issues with stable electrically-charged dark mesons.

Specializing to the case of $N_D = 4$, we then applied our earlier
model-independent lattice results \cite{Appelquist:2014jch} 
to the parameters of stealth dark matter, and obtained 
constraints on the effective Higgs interaction.  
We find that the present LUX bound is able only to mildly 
constrain the Higgs coupling to stealth dark matter for relatively 
light dark baryons.  Even weaker constraints arise when the 
effective Higgs interaction is quadratic in the Yukawa coupling,
which is a natural possibility when the two pairs of dark fermions
are split dominantly by vector-like masses, i.e., $y v \ll \Delta$.

While we have considered many aspects of stealth dark matter,
several avenues warrant further investigation:
\begin{itemize}
\item Chiral symmetry forbids additive renormalization of the fermion masses; we have focused on the regime where the constituent fermion
mass is comparable to the confinement scale 
$M_f \sim \Lambda_D$, since this is best-suited for lattice
simulations, exactly where analytic estimates are least useful. 
It would be interesting to consider a broader range of fermion
masses relative the confinement scale, to understand the relative
scaling of the Higgs interactions.
\item A more precise calculation of the $S$ parameter
is possible  
using lattice simulations for the relevant correlators.
This would allow us to place numerical bounds on the 
parameters of the theory, that could be stronger than the
bounds from the non-observation through direct detection.
\item We would like to unpack $y_{\rm eff}$ [c.f.~Eq.~(\ref{eq:yeff})] 
and obtain constraints on the Yukawa couplings of the model. 
However, this requires translating the fermion masses from the 
lattice regularization into a continuum regularization. 

\item Dark meson production and decay at the LHC is ripe
for exploration.  Dark meson pair production would 
proceed through off-shell EW gauge bosons, $q \bar{q} \ra \Pi^+ \Pi^-$, 
$q \bar{q} \ra \Pi^0 \Pi^0$, and $q \bar{q}' \ra \Pi^\pm \Pi^0$. 
These could have spectacular signals at the LHC\@. 
Neutral mesons decay into fermion pairs and dibosons (explored in other related models in 
\cite{Kilic:2009mi,Kilic:2010et,Fok:2011yc,Buckley:2012ky}). 
For charged dark mesons, with masses in the range 
$m_{\Pi^\pm} \sim 90-180$~GeV, the decay $\Pi^+ \ra \tau^+ \nu_\tau$
dominates, while for masses above this, $\Pi^+ \ra t\bar{b}$ is dominant.  
Charged pion pair production could therefore lead to 
$t \bar{b} b \bar{t}$ signals with the $t \bar{b}$ and $b \bar{t}$
pairs reconstructing to the same mass.  To the best of our knowledge,
this type of resonance search is not being performed at the LHC\@.
\item More insight into the thermal abundance of stealth dark matter,
perhaps using lattice simulations, would help narrow the interesting
mass range that matches cosmological data.
\item Asymmetric production of stealth dark matter seems very
promising, but has several calculational obstacles to overcome
to arrive at a quantitative relationship between the abundance 
and the other parameters of the theory. 
\item If stealth dark matter has an asymmetric abundance, there are
potential limits from neutron star lifetimes 
\cite{McDermott:2011jp,Bramante:2013hn,Bertoni:2013bsa}
though the precise bounds depend sensitively on the equation of state of the 
neutron stars.
\item There are tantalizing signals of a $\gamma$-ray excess between
about $1$-$10$~GeV in the galactic center 
(see for example 
\cite{Hooper:2010mq,Hooper:2011ti,Abazajian:2012pn,Bringmann:2012ez,Daylan:2014rsa}).
A recent analysis \cite{Agrawal:2014oha} suggests that this could 
arise from dark matter up to $300$~GeV\@.  It is intriguing to 
consider the $\gamma$-ray signal spectrum that could arise from a 
symmetric abundance of stealth dark matter with annihilation 
into a multibody final state \cite{Elor:2015tva} with mixtures of 
four or more heavy fermions and multi-gauge bosons
(from $B B^* \ra \Pi \, \Pi \, \ldots \ra$~SM states). 
\end{itemize}

Finally there are broader model-building questions to 
consider.  One is the choice of scales $M_f \sim \Lambda_D$ 
that has been the focus of this work.
This could arise dynamically.  For example, if there are
sufficient flavors in the $SU(N_D)$ gauge theory such that it is
approximately conformal at high energies, then as
the theory is run down through the dark fermion mass scale $M_f$,
the dark fermions integrate out, and confinement sets in at 
$\Lambda_D \sim M_f$. 
This is well known to occur for supersymmetric $SU(N)$ theories 
in the conformal window that flow to confining theories 
once the number of flavors drops below $N_f < 3 N/2$ \cite{Seiberg:1994pq}.
The origin of the vector-like masses of the fermions is also an interesting
model-building puzzle.  However, just as SM fermion masses are
vector-like below the electroweak breaking scale, 
we can imagine dark fermion vector-like masses could be revealed
as arising from dynamics that breaks the 
flavor symmetries of our dark fermions at some higher scale.

\section{Acknowledgments}

We thank S.~Chang, O.~DeWolfe, and D.~B.~Kaplan for many valuable
discussions during the course of this work.  

We thank the Lawrence Livermore National Laboratory (LLNL) Multiprogrammatic and Institutional Computing program for Grand Challenge allocations and time on the LLNL BlueGene/Q 
(rzuseq and vulcan) supercomputer.
We thank LLNL for funding from LDRD~13-ERD-023 
``Illuminating the Dark Universe with PetaFlops Supercomputing''.  
Computing support for this work comes from the 
LLNL Institutional 
Computing Grand Challenge program.  

This work has been supported by the U.~S.~Department of Energy under Grant Nos. DE-SC0008669 and DE-SC0009998 (D.S.), DE-SC0010025 (R.C.B., C.R., E.W.), DE-FG02-92ER40704 (T.A.), DE-SC0011640 (G.D.K.), DE-FG02-00ER41132 (M.I.B.), and Contracts DE-AC52-07NA27344 (LLNL), DE-AC02- 06CH11357 (Argonne Leadership Computing Facility), and by the National Science Foundation under Grant Nos. NSF PHY11-00905 (G.F.), OCI-0749300 (O.W.). Brookhaven National Laboratory is supported by the U.~S.~Department of Energy under contract DE-SC0012704.  S.N.S was supported by the Office of Nuclear Physics in the U.~S.~Department of Energy's Office of Science under Contract DE-AC02-05CH11231.

\appendix

\section{\textbf{Weak Currents}}
\label{sec:weakcurrents}

We examine the dark fermion contributions to the electroweak currents.
In the gauge eigenstate basis, the currents are 
\begin{eqnarray}
j^\mu_{+} &=& -\frac{1}{\sqrt{2}} 
\left(   {F_1^u}^\dagger \bar{\sigma}^\mu F_1^d
       + {F_2^u}^\dagger \bar{\sigma}^\mu F_2^d \right) \\
j^\mu_{-} &=& -\frac{1}{\sqrt{2}} 
\left(   {F_1^d}^\dagger \bar{\sigma}^\mu F_1^u
       + {F_2^d}^\dagger \bar{\sigma}^\mu F_2^u \right) \\
j^\mu_{3} &=& -\frac{i}{2} 
\sum_{i=1,2} \left(   {F_i^u}^\dagger \bar{\sigma}^\mu F_i^u
                    - {F_i^d}^\dagger \bar{\sigma}^\mu F_i^d \right) \\
j^\mu_{Y} &=& -\frac{i}{2}
\sum_{i=3,4} \left(   {F_i^u}^\dagger \bar{\sigma}^\mu F_i^u
                    - {F_i^d}^\dagger \bar{\sigma}^\mu F_i^d \right) \, .
\end{eqnarray}
In the mass eigenstate basis given by Eqs.~(\ref{eq:vec1})-(\ref{eq:vec4}), 
the currents can be rewritten in terms of the 4-component Dirac fermions
defined by Eqs.~(\ref{eq:diracup}),(\ref{eq:diracdown}).  
After some algebra, one obtains
\begin{widetext}
\begin{eqnarray} \label{eq:currents}
j^\mu_{+} &=& -\frac{1}{\sqrt{2}} 
\Big[ 
\overline{\Psi_1^u} \gamma^\mu 
                    \left( c_1^u c_1^d P_L + c_2^u c_2^d P_R \right) \Psi_1^d +
\overline{\Psi_2^u} \gamma^\mu
                    \left( s_1^u s_1^d P_L + s_2^u s_2^d P_R \right) \Psi_2^d 
\nonumber \\ & &{} \qquad\quad +
\overline{\Psi_1^u} \gamma^\mu
                    \left( c_1^u s_1^d P_L + c_2^u s_2^d P_R \right) \Psi_2^d +
\overline{\Psi_2^u} \gamma^\mu
                    \left( s_1^u c_1^d P_L + s_2^u c_2^d P_R \right) \Psi_1^d 
\Big] \label{eq:jplus} \\
j^\mu_{-} &=& -\frac{1}{\sqrt{2}} 
\Big[
\overline{\Psi_1^d} \gamma^\mu
                    \left( c_1^d c_1^u P_L + c_2^d c_2^u P_R \right) \Psi_1^u +
\overline{\Psi_2^d} \gamma^\mu
                    \left( s_1^d s_1^u P_L + s_2^d s_2^u P_R \right) \Psi_2^u 
\nonumber \\ & &{} \qquad\quad +
\overline{\Psi_1^d} \gamma^\mu
                    \left( c_1^d s_1^u P_L + c_2^d s_2^u P_R \right) \Psi_2^u +
\overline{\Psi_2^d} \gamma^\mu
                    \left( s_1^d c_1^u P_L + s_2^d c_2^u P_R \right) \Psi_1^u 
\Big] \label{eq:jminus} \\
j^\mu_{3} &=& \frac{1}{2} 
\Big[ 
\overline{\Psi_1^u} \gamma^\mu
                    \left( (c_1^u)^2 P_L + (c_2^u)^2 P_R \right) \Psi_1^u +
\overline{\Psi_2^u} \gamma^\mu
                    \left( (s_1^u)^2 P_L + (s_2^u)^2 P_R \right) \Psi_2^u 
\nonumber \\ & &{} \qquad -
\overline{\Psi_1^d} \gamma^\mu
                    \left( (c_1^d)^2 P_L + (c_2^d)^2 P_R \right) \Psi_1^d -
\overline{\Psi_2^d} \gamma^\mu
                    \left( (s_1^d)^2 P_L + (s_2^d)^2 P_R \right) \Psi_2^d 
\nonumber \\ & &{} \qquad +
\overline{\Psi_1^u} \gamma^\mu
                    \left( c_1^u s_1^u P_L + c_2^u s_2^u P_R \right) \Psi_2^u +
\overline{\Psi_2^u} \gamma^\mu
                    \left( s_1^u c_1^u P_L + s_2^u c_2^u P_R \right) \Psi_1^u 
\nonumber \\ & &{} \qquad -
\overline{\Psi_1^d} \gamma^\mu
                    \left( c_1^d s_1^d P_L + c_2^d s_2^d P_R \right) \Psi_2^d -
\overline{\Psi_2^d} \gamma^\mu
                    \left( s_1^d c_1^d P_L + s_2^d c_2^d P_R \right) \Psi_1^d 
\Big] \label{eq:j3} \\
j^\mu_{Y} &=& \frac{1}{2}
\Big[
\overline{\Psi_1^u} \gamma^\mu
                    \left( (s_1^u)^2 P_L + (s_2^u)^2 P_R \right) \Psi_1^u +
\overline{\Psi_2^u} \gamma^\mu 
                    \left( (c_1^u)^2 P_L + (c_2^u)^2 P_R \right) \Psi_2^u 
\nonumber \\ & &{} \qquad -
\overline{\Psi_1^d} \gamma^\mu
                    \left( (s_1^d)^2 P_L + (s_2^d)^2 P_R \right) \Psi_1^d -
\overline{\Psi_2^d} \gamma^\mu
                    \left( (c_1^d)^2 P_L + (c_2^d)^2 P_R \right) \Psi_2^d 
\nonumber \\ & &{} \qquad -
\overline{\Psi_1^u} \gamma^\mu
                    \left( c_1^u s_1^u P_L + c_2^u s_2^u P_R \right) \Psi_2^u -
\overline{\Psi_2^u} \gamma^\mu
                    \left( s_1^u c_1^u P_L + s_2^u c_2^u P_R \right) \Psi_1^u 
\nonumber \\ & &{} \qquad +
\overline{\Psi_1^d} \gamma^\mu
                    \left( c_1^d s_1^d P_L + c_2^d s_2^d P_R \right) \Psi_2^d +
\overline{\Psi_2^d} \gamma^\mu
                    \left( s_1^d c_1^d P_L + s_2^d c_2^d P_R \right) \Psi_1^d 
\Big] \, , \label{eq:jY} 
\end{eqnarray}
\end{widetext}
where $c_i^j \equiv \cos\theta_i^j$, $s_i^j \equiv \sin\theta_i^j$ and
$P_{L,R} = (1 \mp \gamma_5)/2$ are the left- and right-handed projectors.
In general, the dark fermions contribute to both the vector
and axial currents with strengths given by the mixing angles.
It is easy to verify that the electromagnetic current, 
\begin{eqnarray}
j_{\rm em}^\mu &=& j_{3}^\mu + j_{Y}^\mu \nonumber \\
&=& \sum_{i=1,2} \Big[
  Q_u \overline{\Psi^u_i} \gamma^\mu \Psi^u_i
+ Q_d \overline{\Psi^d_i} \gamma^\mu \Psi^d_i
\Big] \, ,
\end{eqnarray}
with $Q_{u,d} = \pm 1/2$, is consistent with a pure vector 
coupling of the dark fermions to the photon independent of 
mass mixing angles.  

Interestingly, if the mass matrices
Eqs.(\ref{eq:upmassmatrix}),(\ref{eq:downmassmatrix})  
are symmetric, i.e., $y_{14}^u = y_{23}^u$ and $y_{14}^d = y_{23}^d$,
then just two mixing angles are required, i.e., 
$\theta_1^u = \theta_2^u$ and $\theta_1^d = \theta_2^d$.
In this case, the mixing angles factor out of the 
left-right gamma matrix structure, leaving all of the 
electroweak currents to be purely vector (with vanishing 
axial current).  This is unlike the Standard Model,
where the $SU(2)_L$ currents are purely $V-A$.
The difference between this model and the Standard Model
is the structure of the dark fermion mass matrices that include 
both vector-like and electroweak symmetry breaking masses.

It is also interesting to calculate the neutral current
\begin{eqnarray}
j_{Z}^\mu &=& j_{3}^\mu - \sin^2\theta_W j_{\rm em}^\mu \, .
\end{eqnarray}
For the neutral baryon state, 
\begin{eqnarray}
\langle B | j_Z^\mu | B \rangle &\simeq& \\
& & 
\!\!\!\!\!\!\!\!\!\!\!\!\!\!\!\!\!\!\!\!\!\!\!\!\!\!\!\!\!\!\!
+ \frac{1}{4} \left( (c_1^u)^2 + (c_2^u)^2 - (c_1^d)^2 - (c_2^d)^2 \right) 
\langle B | \overline{\Psi_1} \gamma^\mu \Psi_1 | B \rangle \nonumber \\
& & 
\!\!\!\!\!\!\!\!\!\!\!\!\!\!\!\!\!\!\!\!\!\!\!\!\!\!\!\!\!\!\!
+ \frac{1}{4} \left( - (c_1^u)^2 + (c_2^u)^2 + (c_1^d)^2 - (c_2^d)^2 \right) 
\langle B | \overline{\Psi_1} \gamma^\mu \gamma^5 \Psi_1 | B \rangle \, . 
\nonumber 
\end{eqnarray}
In the limit of zero momentum exchange ($Q^2 = 0$), 
the vector form factor 
$\langle B | \overline{\Psi_1} \gamma^\mu \Psi_1 | B \rangle$
evaluates to $1$, while the axial-vector form factor 
$\langle B | \overline{\Psi_1} \gamma^\mu \gamma^5 \Psi_1 | B \rangle$
for a scalar baryon vanishes.  
In the presence of an exact custodial SU(2) symmetry, 
which is the focus of this paper, we have $c_i^u = c_i^d$ and the $Z$ coupling vanishes identically at any momentum exchange.

On the other hand, if custodial symmetry is broken, then the lightest neutral baryon acquires 
tree-level couplings to the $Z$ boson.  To illustrate the size of these couplings,
consider taking the dark fermion mass matrices to be 
exactly symmetric ($y_{23} = y_{14}$) but allowing for a small, 
custodial symmetry-violating difference in the Yukawas, 
$y_u = y + \xi$ and $y_d = y - \xi$ where $\xi/y \ll 1$. 
The coefficient of the weak neutral vector current becomes
\begin{widetext}
\begin{eqnarray}
(c_1^u)^2 + (c_2^u)^2 - (c_1^d)^2 - (c_2^d)^2 &\simeq&
  \begin{dcases*} 
  2 \sqrt{2} \frac{\xi}{y} \frac{\Delta}{y v} & \quad \mbox{Linear Case} \\
  \frac{\xi}{y} \frac{(y v)^2}{\Delta^2} & \quad \mbox{Quadratic Case.} 
  \end{dcases*} 
\end{eqnarray}
\end{widetext}
Custodial symmetry violation is therefore restricted
by requiring the coupling of the lightest neutral baryon to the $Z$
boson be small enough to have evaded direct detection.
There are several limits in which this can occur: 
$\xi/y \ll 1$ (any scenario), 
$\Delta/(y v) \ll 1$ (Linear Case), or 
$(y v)/\Delta \ll 1$ (Quadratic Case).
This suggests that modest custodial symmetry violation 
is possible but rather constrained.

\bibliography{SU4}

\end{document}